\documentclass[aps,pra,twocolumn,superscriptaddress,floatfix]{revtex4-1}

\usepackage[utf8]{inputenc}
\usepackage[T1]{fontenc}
\usepackage{graphicx}
\usepackage{amsmath}
\usepackage{amssymb}
\usepackage{amsthm}
\usepackage{bm}
\usepackage{caption}
\usepackage{subcaption}
\usepackage{url}

\usepackage{booktabs}
\usepackage{multirow}
\usepackage[
  colorlinks  = true,
  citecolor   = cyan,
  linkcolor   = red
]{hyperref}
\usepackage{cleveref}
\usepackage{physics}
\usepackage{natbib}
\usepackage{enumitem}

\usepackage[table]{xcolor}
\usepackage{colortbl}

\definecolor{gateorange}{RGB}{255,140,0}
\definecolor{gatedeepblue}{RGB}{25,25,112}
\definecolor{gategreen}{RGB}{60,179,113}
\definecolor{gateteal}{RGB}{0,128,128}
\definecolor{lilac}{RGB}{195,185,228}
\definecolor{lilacD}{RGB}{120,100,170}

\usepackage{tikz}
\usepackage{pgfplots}
\pgfplotsset{compat=1.18}
\usetikzlibrary{fit, backgrounds, calc}
\usepackage{quantikz}

\theoremstyle{definition}
\newtheorem{theorem}{Theorem}
\newtheorem{lemma}{Lemma}

\newtheorem{proposition}{Proposition}

\newtheorem*{theorem*}{Theorem}
\newtheorem*{corollary*}{Corollary}
\newtheorem*{lemma*}{Lemma}
\newtheorem*{definition*}{Definition}
\newtheorem*{remark*}{Remark}
\newtheorem*{proposition*}{Proposition}
\theoremstyle{remark}

\allowdisplaybreaks

\begin{document}

\title{Norm-Aware Product Formulas for Hamiltonian Simulations} 

\author{Carlos A. Piñeiro Moreno}
\affiliation{Institute of Inorganic Chemistry, Optical Materials: Computational Methods, Leibniz University Hannover, Callinstr. 9, 30167 Hannover, Germany}
\affiliation{Cluster of Exzellence PhoenixD, Leibniz University Hannover, Welfengarten 1A, 30167 Hannover, Germany}

\author{Pascal Krause}
\affiliation{Institute of Inorganic Chemistry, Optical Materials: Computational Methods, Leibniz University Hannover, Callinstr. 9, 30167 Hannover, Germany}
\affiliation{Cluster of Exzellence PhoenixD, Leibniz University Hannover, Welfengarten 1A, 30167 Hannover, Germany}

\author{Annika Bande}
\email{annika.bande@aci.uni-hannover.de}
\affiliation{Institute of Inorganic Chemistry, Optical Materials: Computational Methods, Leibniz University Hannover, Callinstr. 9, 30167 Hannover, Germany}
\affiliation{Cluster of Exzellence PhoenixD, Leibniz University Hannover, Welfengarten 1A, 30167 Hannover, Germany}
\affiliation{Theory of Electron Dynamics and Spectroscopy, Helmholtz-Zentrum Berlin für Materialien und Energie GmbH, Hahn-Meitner-Platz 1, 14109 Berlin, Germany}

\date{\today}

\begin{abstract}
With the target of efficiently solving the quantum-mechanical Schrödinger equation on future quantum computers, first- and second-order product formulas for the time-evolution 
operator are proposed, designed to exploit hierarchies in the operator 
norms of the individual Hamiltonian components. Inspired by the THRIFT 
product formula, these decomposition schemes achieve a suppression of the 
leading error term by a factor of $\alpha<1$ with respect to the 
standard Lie--Trotter--Suzuki splitting of the same order, where 
$\alpha$ quantifies the norm ratio between the dominant and 
subdominant parts of the Hamiltonian. Unlike THRIFT, the proposed 
formulas retain the general product-formula structure, making them 
applicable to Hamiltonians whose terms act on an arbitrary number of 
qubits. The advantages of this approach are demonstrated through 
numerical experiments on engineered two-dimensional spin-$\frac{1}{2}$ XY model 
systems and isolated molecules within the 
double-factorization framework. 
\end{abstract}

\maketitle

\section{Introduction}

In recent decades, quantum computing (QC) has attracted significant attention from the scientific community. Numerous efforts have been undertaken to transform a purely theoretical idea into a practical computational paradigm~\cite{Feynman1982}. Hence, QC is widely regarded as a promising approach that may offer significant computational advantages across many research areas~\cite{Montanaro2016}.  Moreover, quantum computing has emerged as a technology to address the broad set of challenges that such problems represent for classical computing~\cite{Cao2019QuantumChemistry}. Particularly in quantum chemistry QC poses to be a powerful tool for ultimately finding accurate solutions for molecular electronic structure calculations and the simulations of quantum dynamical processes~\cite{Roncevic2026HalfMobius, Ollitrault2021MolecularQD}.

In this context, quantum simulation constitutes a key application of QC. In quantum simulation, the time evolution of a closed quantum system is characterized by its propagator. For problems that can be described by a time-independent Hamiltonian $H$, this corresponds to implementing the unitary time-evolution operator 

\begin{equation}\label{eq:propagator}
    U(t)=e^{-iHt} .
\end{equation}

The central objective of quantum simulation is therefore to construct an approximation $\tilde{U}(t)$ such that:

\begin{equation}\label{eq: }
    \|U(t)-\tilde{U}(t)\| < \epsilon,
\end{equation}
where $\|\cdot\|$ denotes the operator norm and $\epsilon$ represents the desired target error threshold, while minimizing the required computational resources. There is a wide range of algorithms in the literature for this purpose, all with their specific advantages and disadvantages. Their use depends largely on the characteristics of the addressed problem as well as on the available resources~\cite{McArdle2020_QuantumComputationalChemistry}. The probably most advanced and efficient technique known to date is Quantum Singular Value Transformation, as it is exhibiting the best scaling with respect to the total evolution time $t$ and the targeted $\epsilon$~\cite{LowChuang2017OptimalQSP,Gilyen2019QSVT}. However, for Noisy Intermediate-Scale Quantum Computers, this approach is unfeasible due to the amount of resources required for its implementation in real quantum hardware~\cite{Martyn2021_GrandUnificationQuantumAlgorithms}.

Consequently, attention turns to approaches that are more compatible with current hardware capabilities. In this regard, product formulas remain a family of methods representing the simplest digital approach to approximating the time-evolution operator. These methods take advantage of the fact that the system Hamiltonian is in the majority of cases given as a sum of terms, $H=\sum_k H_k$, each of which can be efficiently implemented as elementary unitary operation at the quantum circuit level. The resulting approximation to the time-evolution operator takes the form~\cite{TrotterCommutatorScaling2021}:
\begin{equation}\label{eq:product_formula}
\tilde{U}(t)=\prod_j\prod_k e^{-ita_{(j,k)}H_{\pi_j(k)}} .
\end{equation} 

Depending on the choice of the real coefficients $a_{(j,k)}$ and the permutation index function $\pi_j(k)$, the product formula achieves a certain approximation order in the time $t$. That is, the norm difference between the exact propagator~\eqref{eq:propagator} and its product formula approximation~\eqref{eq:product_formula} is a function of a power of $t$ and that constitutes the approximation order. As shown in Ref.~\cite{TrotterCommutatorScaling2021}, the error associated with the product-formula approximation can be characterized in terms of the norms of nested commutators of the Hamiltonian components, such as $\|[H_1,H_2]\|$, $\|[H_1,[H_2,H_3]]\|$, and higher-order terms, which determine the scaling of the approximation error. In this sense, nested commutators involving operators with larger norms are expected to yield larger contributions to the total error.

Despite the fact that many commonly used product formulas, such as
Lie--Trotter--Suzuki formulas~\cite{Suzuki1990}, Forest--Ruth~\cite{ForestRuth1990},
and Yoshida integrators~\cite{Yoshida1990}, do not explicitly exploit
norm hierarchies among the summand operators, there have been several formulations addressing this feature. Such Hamiltonians can have, for example, the general form:

\begin{equation}\label{eq:thrift_hamiltonian}
 H = H_0 + \alpha H^{(1)} \quad \text{with} \quad H^{(1)} = \sum_{k=1}^{L} H_k^{(1)}\quad ,
\end{equation}
where $H_0$ and $H_k^{(1)}$ are operators with norms of comparable magnitude, $\alpha$ is a real parameter of small magnitude, and $L$ denotes the total number of operator terms composing the operator $H^{(1)}$. For instance, Omelyan \textit{et al.}~\cite{Omelyan2002Optimized} optimized the fourth-order Forest--Ruth product formula for Hamiltonians with this form for $L=1$, achieving a significantly improved error scaling compared to the original formulation. A recently introduced product formula, called THRIFT~\cite{THRIFT2025}, builds upon this idea by explicitly considering these kinds of Hamiltonians, achieving a suppression of the leading error term by a factor of $\alpha$ with respect to the Lie--Trotter--Suzuki (LTS) decomposition of the same order. However, the systems considered for benchmarking the THRIFT approach so far involved Hamiltonians that are intentionally engineered to satisfy the dominance condition~\cite{THRIFT2025}. Moreover, the THRIFT product formula requires implementing operators of the form $e^{-i(H_0+\alpha H^{(1)}_k)t}$ -- rather than a more deliberte form as of Eq. \eqref{eq:product_formula} --, whose specific structure determines their realization at the circuit level. For the systems analyzed in reference~\cite{THRIFT2025}, this operator corresponds to gates acting on at most two qubits and can therefore be implemented efficiently. 

Throughout this work, product formulas building on the THRIFT approach are proposed that explicitly assume a Hamiltonian of the type \eqref{eq:thrift_hamiltonian}, while maintaining the general form~\eqref{eq:product_formula}, and aiming to remain competitive over a wide range of the parameter $\alpha$. First- and second-order product formulas are considered, as they provide the best balance between accuracy and resource requirements and are among the most commonly used in practical applications.

The notation adopted for the product formulas follows a general form which will be used consistently throughout the remainder of the paper: The approximation $\tilde{U} \rightarrow U_{\mathrm{Name}}^{[\mathrm{Order}]}$. [Order] denotes the order of the method and [Name] denotes the abbreviation or name of the corresponding decomposition. 

The paper is organized such that section~\ref{sec:methods} introduces the theoretical framework and presents the methodology/product formulas employed throughout this work. Section~\ref{sec:XY model} presents numerical experiments on the two-dimensional spin-$\frac{1}{2}$ XY model, where the developed product formulas are compared with the performance of the THRIFT and LTS as well. Section~\ref{sec:molecular} exploits a similar comparison for molecular Hamiltonians in computing the real time propagation of quantum states under the molecular Hamiltonian. Finally, Section~\ref{sec:conclusions} summarizes the main findings of this work.

\section{Methods}\label{sec:methods}

\subsection{From THRIFT to general Product Formulas}\label{sec:main_motivation}

The key idea is to derive generalized classes of product formulas~\eqref{eq:product_formula} applicable to a broader range of problems than those encompassed by the THRIFT product formulas (see~\cite{THRIFT2025}), while retaining the favorable $\alpha$-scaling of the product-formula error characteristic of the THRIFT construction. To this end, the THRIFT first-order decomposition and the error arising from it is shown first. 

The following theorem taken from Ref.~\cite{THRIFT2025} is recalled for completeness.

\begin{theorem*}[THRIFT first-order decomposition~\cite{THRIFT2025}]
Considering the exact unitary time evolution \eqref{eq:propagator} and the Hamiltonian ~\eqref{eq:thrift_hamiltonian},
the \emph{THRIFT approximation} is defined as:
\begin{align}\label{eq:thrift_approx}
  U_{\text{THRIFT}}^{[1]}(t)
  &= e^{-i H_0 t}
    \prod_{k=1}^{L}
    \Bigl(
      e^{i H_0 t}\,
      e^{-i (H_0 + \alpha H_k^{(1)}) t}
    \Bigr).
\end{align}
Then the operator norm of the difference among exact and approximate THRIFT term becomes $\bigl\| U(t) - U_{\text{THRIFT}}^{[1]}(t) \bigr\| = \mathcal{O}(\alpha^2 t^2)$
for sufficiently small $t$.
\end{theorem*}

\medskip

As can be seen from the product of $U_{\text{THRIFT}}^{[1]}(t)$, the operator $e^{-i(H_0+\alpha H_k^{(1)})t}$ must be chosen properly to admit an efficient implementation. Consequently, the practical performance of THRIFT relies on the existence of a resource-efficient synthesis of the exponentials of the grouped operators. Such syntheses are available for operators acting on one or two qubits through Cartan decompositions and related techniques~\cite{Helgason1978,Kokcu2022Cartan}. This requirement is naturally satisfied for the benchmark systems considered in Ref.~\cite{THRIFT2025}, where the Hamiltonians consist exclusively of one- and two-qubit interactions. However, no comparable general-purpose synthesis is currently known for exponentials of arbitrary $k$-qubit operators ($k>2$). As a result, although the THRIFT construction is formally applicable to more general Hamiltonians, its practical advantage is presently limited to settings in which these grouped exponentials can be implemented efficiently.
 
The following analysis overcomes this limitation and ultimately introduces the first-order splitting proposed in this work. The construction starts with the operator $e^{-iH_0t}$, followed by a generic decomposition block representing the product in Eq.~\eqref{eq:thrift_approx}. The last block may in turn be approximated using the \textit{second-order symmetric} Lie--Trotter--Suzuki (Strang) decomposition~\cite{TrotterCommutatorScaling2021}, as given in the following equation. The maximal error arising from this approximation becomes:

\begin{widetext}
\begin{align}\label{eq:strang_approx_error}
\Bigg\|
  \prod_{k=1}^{L} e^{i H_0 t} e^{-i (H_0 + \alpha H_k^{(1)})t}
  - 
      \prod_{k=1}^{L} e^{i H_0 t} \bigl(
      e^{-i H_0 t/2} e^{-i \alpha H_k^{(1)} t} e^{-i H_0 t/2}
    \bigr)
\Bigg\|
      &\le \sum_{k=1}^{L} 
   \Big\|
     e^{-i (H_0 + \alpha H_k^{(1)})t}- 
     e^{-i H_0 t/2} e^{-i \alpha H_k^{(1)} t} e^{-i H_0 t/2}
   \Big\| \nonumber\\ 
      &\le C t^3 = \mathcal{O}(|\alpha|t^3).
\end{align}
\end{widetext}

Here the submultiplicative property of the operator norm and the triangle inequality have been used to bound the total error by the sum of the individual block errors, as well as the characteristic $t$-scaling of the Strang decomposition (see below for further details). A worst-case estimation of the parameter $C$ could be [cite or say where/how you prove it]:
\begin{align}\label{eq:t_scaling_partial_strang}
    C=\frac{|\alpha|L}{24}\,
    \max_{k}
    \Bigl(
      2|\alpha| \bigl\|[H_k^{(1)},[H_k^{(1)},H_0]]\bigr\| \nonumber\\+
      \bigl\|[H_0,[H_0,H_k^{(1)}]]\bigr\|
    \Bigr).
\end{align}
$C$ is proportional to $|\alpha|$ as well as the number $L$ of $H^{(1)}$ operators. This $\mathcal{O}(|\alpha|t^3)$ error is of higher order than the
$\mathcal{O}(\alpha^2t^2)$ error of the first-order THRIFT formula, so Strang's approximation
does not degrade the overall accuracy at the limit $t\rightarrow0$, while allowing the general product formula form \eqref{eq:product_formula} to be recovered.

\subsection{First-Order Product Formulas}

Although not fully rigorous, the argument presented in the last section suggests a clear path forward:
If the Strang splitting of $e^{-i(H_0+\alpha H_k^{(1)})t}$ introduces only
a higher-order error, one can build a practically implementable product
formula that still inherits the $\alpha$-suppression relative to the first-order
Lie--Trotter--Suzuki error. Make this idea precise is the purpose of this section (see Theorem \ref{thm:ps_first_order} later). For clarity and for use throughout the subsequent sections of this paper, the following lemma is introduced:

\begin{lemma}[Second-Order $\lambda$ Decomposition]\label{lem:lambda_decomp}
Let $H = H_0 + \alpha H^{(1)}$ (the Hamiltonian has only two terms in its decomposition) with $\alpha \in \mathbb{R}$. Then, for any real $\lambda \in \mathbb{R}$, the following product formula provides a second-order approximation for small $t$:
\begin{align}\label{eq:lambda_decomp}
e^{-iHt}
&=
e^{-i \lambda H_0 t}
\, e^{-i \alpha H^{(1)} t/2}
\, e^{-i (1-2\lambda) H_0 t}
\, e^{-i \alpha H^{(1)} t/2}
\, \notag \\ 
&\times e^{-i \lambda H_0 t}
+ \mathcal{O}(\alpha t^{3}).
\end{align}
\end{lemma}

Moreover, the leading local error at order $\mathcal{O}(\alpha t^{3})$ is
\begin{equation}\label{eq:lambda_error}
  -i\alpha t^{3}\!\left(
    \gamma_1\,[H^{(1)},[H^{(1)},H_0]]
    +
    \gamma_2\,[H_0,[H_0,H^{(1)}]]
  \right),
\end{equation}
with coefficients
\begin{equation}\label{eq:gamma_coeffs}
  \gamma_1 = -\alpha \!\left(\frac{1 - 6\lambda}{24}\right),
  \qquad
  \gamma_2 = -\frac{1}{12} + \frac{\lambda}{2} - \frac{\lambda^2}{2}.
\end{equation}

\noindent\textbf{Proof.}  See Appendix ~\ref{app:lambda_proof}.

\medskip
\noindent\textbf{Remarks.}
\begin{enumerate}[label=(\roman*)]
  \item \textbf{Optimal $\lambda$ for small $\alpha$.}
Following the argument of Omelyan \textit{et al.}~\cite{Omelyan2003Symplectic}, the
optimal parameter $\tilde{\lambda}$ is defined as the value of $\lambda$
that minimizes the quadratic sum of the coefficients $\gamma_1$ and $\gamma_2$, i.e., the quantity (the explicit dependency of $\gamma_1$ on $\alpha$ is omitted):

\begin{equation}
   \tilde{\lambda}(\alpha)= \arg\min_\lambda\left (\gamma_1^2(\lambda,\alpha) + \gamma_2^2(\lambda ) \right) \quad. 
\end{equation}
In the regime of $\alpha$ being small, the contribution of $\gamma_1$ can be neglected, since $\gamma_1 \propto \alpha$, such that the optimization is effectively governed by $\gamma_2$. The task then reduces to minimizing $|\gamma_2 (\lambda)|$, yielding
$\tilde{\lambda}(\alpha \to 0) \to 0.211325$ (the second real zero of the polynomial is discussed in Appendix~\ref{app:lambda_proof}),
in agreement with the result reported in~\cite{barthel2020optimized},
although obtained there via a different derivation. Even if $\alpha$ was not small, e.g. $\alpha \to 1$,
the balance between terms only changes slightly, leading to a shift in the
optimal parameter to $\tilde{\lambda}(\alpha \to 1) \to 0.193183$, consistent with
the value reported earlier by Omelyan \textit{et al.}~\cite{Omelyan2003Symplectic} (see
Appendix~\ref{app:lambda_proof}).

  \item \textbf{Recovery of the Strang splitting.}
        Setting $\lambda = 1/2$ in Eq.~\eqref{eq:lambda_decomp} recovers
        the well-known symmetric product formula for two operators
        (Eq.~(L5) of Ref.~\cite{TrotterCommutatorScaling2021}).
\end{enumerate}

Using Lemma~\ref{lem:lambda_decomp} with $\lambda = 1/2$, the symmetric
splitting reads:
\begin{equation}\label{eq:symmetric_strang}
  e^{-i H_0 t/2}
  \, e^{-i \alpha H^{(1)} t}
  \, e^{-i H_0 t/2},
\end{equation}
where the two central operators  $e^{-i \alpha H^{(1)} t/2}$ have been merged into a single one. Then, the operator $e^{-i \alpha H^{(1)} t}$ can be approximated by its first-order Lie--Trotter--Suzuki product formula.  Because this approximation only affects the third-order error, the
overall scaling is preserved, leading to the following main theorem:

\begin{theorem}[First-Order Partial--Strang (PS) Decomposition]
\label{thm:ps_first_order}
Let $H$ be defined as in Eq.~\eqref{eq:thrift_hamiltonian}, with $\alpha$ small in magnitude and $U(t)$ defined as in  Eq.~\eqref{eq:propagator}.
Then the \emph{Partial--Strang} decomposition
\begin{equation}\label{eq:ps_formula}
  U_{\text{PS}}^{[1]}(t)
  = e^{-i H_0 t/2}
    \left(
      \prod_{k=1}^{L} e^{-i \alpha H_k^{(1)} t}
    \right)
    e^{-i H_0 t/2}
\end{equation}
satisfies
$\bigl\| U(t) - U_{\text{PS}}^{[1]}(t) \bigr\| = \mathcal{O}(\alpha^2 t^2)$.
\end{theorem}

\noindent\textbf{Proof.} In the first two steps the scaling in $t$ of the errors arising from the two approximations are calculated, in step 3 compared.

\noindent\textit{Step 1: Strang approximation of the joint propagator.}
Using Eq.~\eqref{eq:symmetric_strang} in computing the norm of the difference from the propagator of Eq. \eqref{eq:thrift_hamiltonian} gives
\begin{align}\label{eq:strang_local_error}
  &\Bigl\|
    e^{-i (H_0 + \alpha H^{(1)}) t}
    -
    e^{-i H_0 t/2}\,e^{-i \alpha H^{(1)} t}\,e^{-i H_0 t/2}
  \Bigr\|
  \nonumber\\
  & \le
  \frac{|\alpha| t^3}{24}
  \Bigl(2
    |\alpha| \bigl\| [H^{(1)},[H^{(1)},H_0]] \bigr\|
    + \bigl\| [H_0,[H_0,H^{(1)}]] \bigr\|
  \Bigr)
  \nonumber\\
  & = 	\mathcal{O}(|\alpha| t^3).
\end{align}

\noindent\textit{Step 2: First-order Trotter--Suzuki decomposition of
$e^{-i\alpha H^{(1)} t}$.}`
\begin{align}\label{eq:trotter_alpha_error}
  &\Bigl\|
    e^{-i \alpha H^{(1)} t}
    - \prod_{k=1}^{L} e^{-i \alpha H_k^{(1)} t}
  \Bigr\|
  \nonumber\\
  &\quad \le
  \frac{\alpha^2 t^2}{2}
  \sum_{1 \le j < k \le L}
  \bigl\| [H_j^{(1)}, H_k^{(1)}] \bigr\|
  = \mathcal{O}(\alpha^2 t^2).
\end{align}

\noindent\textit{Step 3: Triangle inequality.}\label{Step_3}
Combining \eqref{eq:strang_local_error} and \eqref{eq:trotter_alpha_error} leads to the error of the Partial Strang approximation as:
\begin{equation}\label{eq:ps_single_step}
  \bigl\| U(t) - U_{\text{PS}}^{[1]}(t) \bigr\|
  \le \mathcal{O}(\alpha^2 t^2) + \mathcal{O}(|\alpha| t^3)
  = \mathcal{O}(\alpha^2 t^2),
\end{equation}
where the dominant contribution in the limit $t\rightarrow0$ is
$\mathcal{O}(\alpha^2 t^2)$, since $\mathcal{O}(|\alpha|t^3)$ vanishes faster as $t\rightarrow0$.

\medskip
\noindent\textbf{Remarks.}
\begin{enumerate}[label=(\roman*)]

  \item \textbf{Gate overhead.}
   The PS splitting~\eqref{eq:ps_formula} requires only one additional exponential factor of $e^{-i H_0 t/2}$ compared to the first-order Lie--Trotter--Suzuki decomposition $U_{\text{LTS}}^{[1]}$, that is:
   
   \begin{equation}\label{eq:first_trotter}
       U_{LTS}^{[1]} (t)=e^{-iH_0t}\prod_{k=1}^{L} e^{-i \alpha H_k^{(1)} t} \quad,
   \end{equation}

while achieving a suppression of the leading error term by a factor of $\alpha$ with respect to $U_{\text{LTS}}^{[1]}$, since the error of $U_{\text{LTS}}^{[1]}$ scales as $\mathcal{O}(\alpha t^2)$ (linear in $\alpha$). The additional resources required to implement PS depend on the specific structure of the operator $H_0$. In the following section it is shown that for the two-dimensional spin-$\frac{1}{2}$ XY model, the extra layer consists solely of single-qubit operations (more precisely, phase rotations on individual qubits). For electronic Hamiltonians, however, the situation is more involved.

 \item \textbf{Different starting $\lambda$ value.} Having used $\lambda = 1/2$ at the starting point of the derivation, an analogous construction for $\lambda \neq 1/2$ can achieve an error reduction by a factor of $\alpha/2$ relative to the Lie--Trotter--Suzuki error. However, this comes at the cost of a larger number of exponentials, making the choice $\lambda = 1/2$ the most efficient starting point for Theorem~\ref{thm:ps_first_order} in terms of the total number of exponentials to be executed. Nevertheless, for completeness and for later use in this work, we define a general form of the PS approximation for an arbitrary $\lambda$, which we refer to as the \textit{$\lambda$-approximation}.

    \begin{align}\label{eq:ps_first_order_general}
    U_\lambda^{[1]}(t) &= e^{-i \lambda H_0 t}
\, \left(\prod_{k=1}^{L} e^{-i \alpha H_k^{(1)} t/2} \right)
\, e^{-i (1-2\lambda) H_0 t} \notag  \\ 
&\times \, \left(\prod_{k=1}^{L} e^{-i \alpha H_k^{(1)} t/2} \right)
\, e^{-i \lambda H_0 t}.
\end{align}
Although Eq.~\eqref{eq:ps_first_order_general} defines a broader one-parameter family of first-order product formulas, it is of limited practical interest. As shown in the following section, second-order product formulas can be constructed using the same, or even fewer, individual exponential operators while achieving a higher order of accuracy. 

\item \textbf{Higher-Order Partial--Strang Decomposition}

    Techniques for extending first-order product formulas to
    higher orders are widely known. Namely, the general formula for these extensions follows the pattern 
    
    \begin{equation}\label{eq:ps_kth_order_general}
         U^{[k]}_{\text{PS}}(t)
      = \prod_{i}
        U_{\text{PS}}^{[1]}\!\left( u_i\,t /2\right)
        \left(U_{\text{PS}}^{[1]}\!\left(- u_i\,t /2\right) \right)^{\dagger}.
    \end{equation}
    
    where $u_i$ are real parameters \cite{Suzuki1992}. Unfortunately, this product formula, unlike that of the first order, does not scale as $\alpha^2$ in the limit $t\rightarrow 0$. To achieve that, the error of approximating THRIFT by PS should be at least one order greater than the scaling of the PS itself, which only happens for the first order (see Eq.~\eqref{eq:strang_approx_error}). 
    
    The generalization for an arbitrary order $k$ of the first-order PS splitting implements exactly $k$ more exponentials of the form $e^{-iaH_0}$ ($a\in\mathbb{R}$) than the $k$th-order Trotter--Suzuki decomposition. In the specific case of second order, Eq.~\eqref{eq:ps_kth_order_general} is reduced to:
    
    \begin{align}\label{eq:ps_second_order_expanded}
      U^{[2]}_{\text{PS}}(t)
      &= e^{-iH_0t/4}
        \Biggl(
           \prod_{k=1}^{L} e^{-i \alpha H_k^{(1)} t/2}
        \Biggr)
        e^{-iH_0t/2}
        \nonumber\\
      &\quad \times
        \Biggl(
           \prod_{k=L}^{1} e^{-i \alpha H_k^{(1)} t/2}
        \Biggr)
        e^{-iH_0t/4}
    \end{align}

    \end{enumerate}

    To obtain a decomposition of the type $\tilde{U}$ of 
    Eq.~\eqref{eq:product_formula} that achieves a suppression of the 
    leading error term by a factor of $\alpha$ with respect to the 
    second-order Lie--Trotter--Suzuki decomposition, a different approach 
    must be followed. The next section addresses this problem.
    
\subsection{Second-Order Product Formulas}
\label{sec:ps_second_order}

Here, the same idea developed above is extended to second order by substituting each $e^{-i \alpha H^{(1)} t/2}$ factor in Eq.~\eqref{eq:lambda_decomp} with the second-order symmetric LTS product generalized to multiple operators, which is given as:

\begin{align}\label{eq:second_order_alpha}
  e^{-it\alpha\sum_{k=1}^LH^{(1)}_k}&=
  \prod_{k=1}^{L} e^{-i \alpha H_k^{(1)} t/4}
  \prod_{k=L}^{1} e^{-i \alpha H_k^{(1)} t/4} \notag \\
  &+ \mathcal{O}\!\left(\alpha^{3} t^{3}\right).
\end{align}

Inserting~\eqref{eq:second_order_alpha} into~\eqref{eq:lambda_decomp} gives:

\begin{widetext}
\begin{align*}
e^{-i \left(H_0 + \alpha \sum_{k=1}^{L} H_k^{(1)}\right)t}
&=
e^{-i \lambda H_0 t}
\Biggl(
    \prod_{k=1}^{L}
        e^{-i \alpha H_k^{(1)} t/4}
    \prod_{k=L}^{1}
        e^{-i \alpha H_k^{(1)} t/4}
\Biggr)
e^{-i (1-2\lambda) H_0 t}
\Biggl(
    \prod_{k=1}^{L}
        e^{-i \alpha H_k^{(1)} t/4}
    \prod_{k=L}^{1}
        e^{-i \alpha H_k^{(1)} t/4}
\Biggr)
e^{-i \lambda H_0 t}
\nonumber\\
&\quad
+\mathcal{O}(\alpha t^{3})
+\mathcal{O}(\alpha^{3}t^{3}).
\end{align*}
\end{widetext}

The two error sources are:
\begin{enumerate}[label=(\roman*)]
  \item $\mathcal{O}(\alpha t^3)$ from the $\lambda$ decomposition
        (Lemma~\ref{lem:lambda_decomp}), controlled by $\gamma_1$ and $\gamma_2$; namely Eqs.~\eqref{eq:lambda_decomp}-\eqref{eq:gamma_coeffs}.
  \item $\mathcal{O}(\alpha^3 t^3)$ from approximation~\eqref{eq:second_order_alpha}.
\end{enumerate}
Choosing $\lambda $ such that $\gamma_2 = 0$ makes the
dominant error proportional to $\alpha^2$ (Eq.\eqref{eq:lambda_error} and Eq.\eqref{eq:gamma_coeffs}), achieves the desired reduction. Putting all of these pieces together, one can construct a second-order product formula with an error scaling proportional to $\alpha^2$, as is done in

\begin{proposition}\label{prop:ps_second_order}
Let $\tilde{\lambda} = \tfrac{1}{2} - \tfrac{\sqrt{3}}{6} \approx 0.211325$, which was discussed in Lemma~\ref{lem:lambda_decomp} to be the optimal $\lambda$ for small $\alpha$. Then the decomposition
\begin{align}\label{eq:ps_second_order_formula}
  & U_{\tilde{\lambda}}^{[2]}(t)
  = e^{-i \tilde\lambda H_0 t}
  \Biggl(
    \prod_{k=1}^{L} e^{-i \alpha H_k^{(1)} t/4}
    \prod_{k=L}^{1} e^{-i \alpha H_k^{(1)} t/4}
  \Biggr)
  \nonumber\\
  & \times  e^{-i (1-2\tilde\lambda) H_0 t}
  \Biggl(
    \prod_{k=1}^{L} e^{-i \alpha H_k^{(1)} t/4}
    \prod_{k=L}^{1} e^{-i \alpha H_k^{(1)} t/4}
  \Biggr) \nonumber\\
  & \times
  e^{-i \tilde\lambda H_0 t}
\end{align}
satisfies
$\bigl\| U(t) - U_{\tilde{\lambda}}^{[2]}(t) \bigr\| = \mathcal{O}(\alpha^2 t^{3})$,
thereby achieving a suppression of the leading $\mathcal{O}(\alpha t^{3})$
error term compared to the standard second-order
Lie--Trotter--Suzuki decomposition.
\end{proposition}

For comparison, the generalization of the product formula~\eqref{eq:lambda_decomp} proposed by Ostmeyer~\cite{Ostmeyer2023OptimisedTrotter}
to $L+1$ operators reads :
\begin{align}\label{eq:omelyan_general}
  &U_{\text{OG}}^{[2]}(t)
  = e^{-i\tilde{\lambda} H_0 t}
  \!\left(
    \prod_{k=1}^{L} e^{-i\tilde{\lambda} \alpha H_k^{(1)} t}
    \prod_{k=L}^{1} e^{-i(1-2\tilde{\lambda}) \alpha H_k^{(1)} t /2}
  \right)
  \nonumber\\
  &\times e^{-i(1-2\tilde{\lambda})H_0 t}
  \!\left(
    \prod_{k=1}^{L} e^{-i(1-2\tilde{\lambda}) \alpha H_k^{(1)} t/2}
    \prod_{k=L}^{1} e^{-i\tilde{\lambda} \alpha H_k^{(1)} t}
  \right) \nonumber\\
  & \times e^{-i\tilde{\lambda} H_0 t},
\end{align}
which differs slightly from $U_{\tilde{\lambda}}^{[2]}(t)$ in the divisors within the exponentials. The subsequent numerical experiments indicate that both constructions exhibit the same asymptotic error scaling over a broad range of $\alpha$.

Notably, setting $\lambda = 1/2$ in Ostmeyer’s scheme reduces the decomposition to the standard Lie--Trotter--Suzuki formula of second order \cite{LowChuang2017OptimalQSP}, namely:

\begin{align}\label{eq:second_trotter}
     U_{\text{LTS}}^{[2]} (t)=&e^{-iH_0t/2}\left(\prod_{k=1}^{L} e^{-i \alpha H_k^{(1)} t/2}\prod_{k=L}^{1} e^{-i \alpha H_k^{(1)} t/2} \right)\notag \\
     &\times e^{-iH_0t/2} \quad.
\end{align}

This comparison clarifies why optimized choices of $\lambda$ yield improved second-order approximations: The standard second-order splitting does not cancel the leading error term linear in $\alpha$, whereas the optimized schemes explicitly suppress this contribution. It is important to note that both product formulas, $U_{\text{OG}}^{[2]}$ and $U_{\tilde{\lambda}}^{[2]}$, require the evaluation of $4L+1$ exponential operators, whereas the second- and fourth-order LTS formulas require $2L+1$ and $10L+1$ exponentials, respectively. Therefore, the number of exponentials required by the proposed formulas lies between those of the second- and fourth-order Trotter–Suzuki methods~\cite{Suzuki1992,TrotterCommutatorScaling2021}. On the other hand, in both the local and global regimes, $U^{[1]}_{PS}$ requires only one additional exponential compared with the first-order LTS decomposition $U_{\text{LTS}}^{[1]}$. Clearly, for a sufficiently small time step, second-order product formulas outperform first-order product formulas. Whenever the available resources and the required conditions for the product formulas presented so far are met, their use could considerably reduce the error throughout the simulation, as will be shown in the next section. 


\subsection{Accumulating errors over time}

\noindent Since the error scalings derived above are valid only for short evolution times $t$ (i.e., they are local error bounds), the product formula must be applied iteratively to simulate the evolution up to an arbitrary final time. By combining the triangle inequality with the unitarity of $U_{\text{PS}}^{[1]}$, the total error can be bounded, as the local errors accumulate at most linearly with the number of time steps~\cite{TrotterCommutatorScaling2021}:

A second convention concerns the time argument. We reserve $t$ for the time spanned by a \emph{single} application of a product formula, and $\Delta t$ for the duration
of one step of an \emph{iterated} scheme: when the evolution up to a total time $t$ is carried out in $N$ steps, each step has duration $\Delta t \equiv t/N$, and the approximate propagator is $[\tilde{U}(\Delta t)]^{N}$. The two symbols therefore denote the same quantity whenever $N=1$. The distinction matters because the error bounds of
Sec.~\ref{sec:methods} are \emph{local}, that is, they refer to one application of the formula and are naturally expressed in terms of its argument; the error actually
observed at the end of a simulation is instead the accumulation of $N$ such contributions. Accordingly, quantities reported as functions of $\Delta t$ at fixed $t$
involve a discretisation into $N$ steps, whereas quantities reported as functions of $t$ alone --- as in Appendix~\ref{app:alpha_Scaling_Analysis} --- involve none.

\section{Results and Discussion}

Two types of testing cases for the different methodologies are used: spin-XY and molecular. There are exact time-evolution solutions available, which serve as a numerical reference against which the error norm is computed. 

All results reported in this work were obtained by exact classical emulation of the product formulas; no quantum hardware or shot-based sampling was involved. Two independent implementations were used, one for each family of Hamiltonians, both written in \texttt{Python} on top of \texttt{NumPy} and \texttt{SciPy}~\cite{harris2020numpy,virtanen2020scipy}. All arithmetic is performed in double precision. The complete source code that reproduces every
figure and table of this work, together with the raw data, is openly available (see the Code
Availability section \ref{sc:Code_availability}).

\subsection{Two-dimensional spin-\texorpdfstring{$\frac{1}{2}$}{1/2} XY model}
\label{sec:XY model}
 
To validate the proposed method, all methodologies presented in Sec. \ref{sec:methods} are applied to the two-dimensional spin-$\frac{1}{2}$ XY model with a homogeneous longitudinal field, considering a $3 \times 3$ spin lattice with open boundary conditions. This reduced system was selected because its exact quantum dynamics can be obtained by explicitly constructing the full Hamiltonian matrix and directly evaluating the time-evolution operator, $e^{-iHt}$, through matrix exponentiation. These simulations serve as a benchmark for comparing the error scaling and efficiency of our reduced approach, quantifying the accuracy and computational advantages of the proposed method. The Hamiltonian of the system studied takes the form:

\begin{equation}\label{eq:XY model}
  H = \sum_m Z_m
      + \sum_{\langle i,j \rangle}
        \bigl( \alpha_x X_i X_j + \alpha_y Y_i Y_j \bigr) \quad.
\end{equation}

All quantities are expressed in dimensionless units, with $(\hbar=1$); time is expressed in the corresponding natural units. Here, the sum over $Z_m$ forms the unperturbed Hamiltonian of Eq. \eqref{eq:thrift_hamiltonian}, i.e. $H_0$. Each term denotes a Pauli-$Z$ operator acting on the $m$-th spin. The term is corresponding to the longitudinal field term and runs over all lattice spins. The double summed terms are the perturbations, denoted $H_k^{(1)}$ in Eq.\eqref{eq:thrift_hamiltonian}. They contain the operators $X_i$ and $Y_i$, which are the standard Pauli-$X$ and Pauli-$Y$ operators acting on the spins $i$ and $j$, respectively, being here by the second sum over $\langle i,j \rangle$ restricted to pairs of nearest-neighbor spins. These operators represent the spin-flip interactions along the $x$ and $y$ axes, which are typically associated with transverse magnetic fields in the two-dimensional spin-$\frac{1}{2}$ XY model. The coefficients $\alpha_x$ and $\alpha_y$ determine the relative strengths of the interactions. They were also selected in a way to avoid accidental commutativity collapses, meaning that if the interaction terms commute with each other, then the first-order PS decomposition \eqref{eq:ps_formula} collapses into a second-order formula. Appropriate parameters $\alpha_x$ and $\alpha_y$ have to be non-equal, such that for bonds sharing a common site $j$, non-commutation is guaranteed by

        \begin{align*}
          &[\alpha_x X_i X_j + \alpha_y Y_i Y_j,\,
            \alpha_x X_j X_m + \alpha_y Y_j Y_m] 
          \nonumber \\
&= 2 i\, \alpha_x \alpha_y Z_j\Big( X_i Y_m - Y_i X_m \Big)\neq 0
          \nonumber \\ &\quad\text{when } \alpha_x \alpha_y \neq 0.
        \end{align*}

On the other hand, another requirement should be met: $H_0$ must not commute with the interaction terms $H_k^{(1)}$, otherwise the THRIFT decomposition will collapse to the first-order Trotter--Suzuki decomposition. This is shown exemplarily for two sites $i$ and $j$, respectively.

\begin{equation*}
    \bigl[Z_i+Z_j,\,
            \alpha_x X_i X_j + \alpha_y Y_i Y_j\bigr] = 2i(\alpha_x - \alpha_y)(Y_i X_j + X_i Y_j)
\end{equation*}
vanishes if and only if $\alpha_x = \alpha_y$.

Several sets of simulations were explored with the objective to analyze the errors associated with the different Hamiltonian splitting methods presented in Sec.~\ref{sec:methods} obtained against an exact reference result as Frobenius norm $\|U-\tilde{U}\|_{F}$~\cite{golub2013}. For the approximations $\tilde{U}$, each elementary factor of the product formula---the one-site field exponential, the two-site field exponential, and the THRIFT and $\lambda$ blocks---is applied to the computational basis by tensor contraction, so that the approximate propagator of a single Trotter step is assembled explicitly. The optimal parameter $\tilde{\lambda}$ and the power-law fits were obtained with \texttt{scipy.optimize}. Since the Hamiltonian is time-independent, the $N$-step propagator is obtained as the $N$th matrix power of that step. On the other hand, the exact reference $U=e^{-iHt}$ is computed from the dense matrix exponential (\texttt{scipy.linalg.expm}).

In the first group of simulations, the parameters were chosen such that $\alpha_x = -\alpha_y = \alpha$, which trivially satisfies the conditions established above. The total evolution time was limited to $t = 1$ a.u., and the error was evaluated after $N = 100$ time steps ($\Delta t = 0.01$ a.u.). 

\begin{figure*}[t]
    \centering

    \begin{subfigure}[b]{0.49\textwidth}
        \centering
        \caption{Propagator error as a function of the interaction strength $\alpha$.}
        \includegraphics[width=\textwidth]{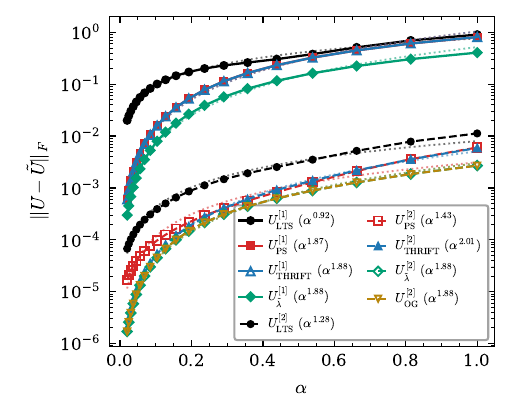}
        \label{fig:error_alpha}
    \end{subfigure}
    \hfill
    \begin{subfigure}[b]{0.49\textwidth}
        \centering
        \caption{Ratio of the errors for selected pairs of methods.}
        \includegraphics[width=\textwidth]{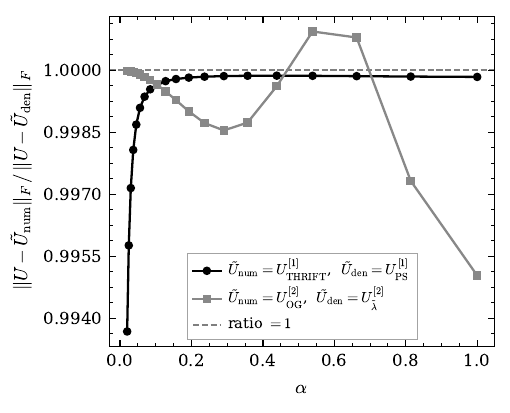}
        \label{fig:ratio_alpha}
    \end{subfigure}

    \caption{Results for the $3\times3$ open-boundary two-dimensional spin-$\frac{1}{2}$ XY model [Eq.~\eqref{eq:XY model}] with $\alpha_x=-\alpha_y=\alpha$, evaluated at $t=1$~a.u.\ after $N=100$ Trotter steps ($\Delta t=0.01$~a.u.). Panel~(\subref{fig:error_alpha}) shows the propagator error $\|U-\tilde U\|_F$, whereas panel~(\subref{fig:ratio_alpha}) shows the ratio of the errors for selected pairs of methods. The methods, line styles, and    power-law fits (dotted) are defined in the main text and the legend.}
    \label{fig:dos_figuras}
\end{figure*}

A particular objective is to investigate the error scaling on the parameter $\alpha$, which shown in Figure~\ref{fig:dos_figuras}\subref{fig:error_alpha} for the diverse methods: The first-order methods compared are $U_{\text{LTS}}^{[1]}$, $U_{\text{PS}}^{[1]}$, $U_{\text{THRIFT}}^{[1]}$, and $U_{\tilde{\lambda}}^{[1]}$, the second-order methods are $U_{\text{LTS}}^{[2]}$, $U_{\text{PS}}^{[2]}$, $U_{\text{THRIFT}}^{[2]}$, $U_{\tilde{\lambda}}^{[2]}$, and $U_{\text{OG}}^{[2]}$ each method with [NAME] with an own color and symbol irrespective of its order. Solid and dashed lines denote first- and second-order methods, respectively, whereas dotted lines represent power-law fits of the form $\propto \alpha^n$, with the fitted exponent $n$ reported in the legend. The fits were performed over $20$ logarithmically spaced values of $\alpha \in [0.02, 1.0]$. All product formulas depending on the parameter $\lambda$ were evaluated using $\tilde{\lambda}$.

All errors in Figure~\ref{fig:dos_figuras}\subref{fig:error_alpha} increase strongly as $\alpha$ increases to 0.1 and then more slowly increase monotonically, while the fitted $\alpha^n$ gives values of $n \in [0.9179;2.0090]$ revealing linear to quadratic behaviors. This is broadly consistent with our theoretical predictions from Sec.~\ref{sec:methods}: except for $U_{\text{LTS}}^{[1]}$ (Eq.~\eqref{eq:first_trotter}, $\mathcal{O}(\alpha t)$, large black points), which exhibits the expected linear scaling in $\alpha$ and $U_{LTS}^{[2]}$ \eqref{eq:second_trotter} (small black points with dashed line) and $U_{\text{PS}}^{[2]}$ (small red squares), which deviate significantly from the quadratic behavior towards lower values, all remaining methods show near-quadratic scaling, consistent with their construction. 

However, as $\alpha \to 1$, the assumptions underlying the theoretical error bounds cease to hold --- in this regime, all operators in the Hamiltonian have considerable norm---, and the error should become indistinguishably each other since there is no $\alpha$ suppression. For the first order methods this breakdown of predictive scaling manifests clearly in the convergence of $U_{LTS}^{[1]}$, $U_{\text{PS}}^{[1]}$, and $U_{\text{THRIFT}}^{[1]}$ (three top lines) toward a common error value as $\alpha \to 1$. Here the perturbative assumptions for small $\alpha$ break down and as a result the structured differences between these methods are washed out and their errors become indistinguishable. On the other hand, for small $\alpha$ the relative performance can be analyzed. Both, $U_{\text{PS}}^{[1]}$ and $U_{\text{THRIFT}}^{[1]}$, remain nearly indistinguishable and outperform $U_{\text{LTS}}^{[1]}$ since both have the same scaling in $t$ and $\alpha$ ($\mathcal{O}(\alpha^2 t)$) with relative deviations between them below $1\%$ throughout the explored range. The $\lambda$ decomposition $U_{\tilde{\lambda}}^{[1]}$ (green diamonds) performs best among the product formulas of first-order, but is, as the overall first-order formulas, outperformed by the product the second-order approximations and is even more expensive to execute than second-order LTS (in terms of the number of exponentials), making it not feasible to consider further.

Among the second-order methods, $U_{\tilde{\lambda}}^{[2]}$ (orange down-triangles), $U_{\text{OG}}^{[2]}$ (cyan open down-triangles), and $U_{\text{THRIFT}}^{[2]}$ (blue triangles) achieve the systematically lowest absolute errors outperforming both $U_{\text{LTS}}^{[2]}$ (black points) and $U_{\text{PS}}^{[2]}$ (red squares) when going to the perturbative regime. When $\alpha \to 1$, $U_{\text{THRIFT}}^{[2]}$ gains in error more quickly, suggesting that the additional variational freedom in $U_{\tilde{\lambda}}^{[2]}$ and $U_{\text{OG}}^{[2]}$ leads to a more favorable error structure even outside the perturbative regime.

The right panel of Figure~\ref{fig:dos_figuras}\subref{fig:ratio_alpha} shows the ratio of Frobenius-norm errors between selected pairs of methods under the same conditions as the simulations in the left panel; a ratio below unity indicates that the numerator method outperforms the denominator in terms of accuracy (lower error). The pairs selected are $U_{\text{THRIFT}}^{[1]}$ and $U_{\text{PS}}^{[1]}$ for first-order methods and $U_{\text{OG}}^{[2]}$ and $U_{\tilde{\lambda}}^{[2]}$ for second-order methods. The selection corresponds to product formulas with similar performance over the entire range of $\alpha$ shown in panel~\ref{fig:dos_figuras}\subref{fig:error_alpha}. Turning to the quotient between the first-order methods (black filled circles), one can see that both methods perform almost identically over a broad range of $\alpha$ values. However, this behavior breaks down as $\alpha \rightarrow 0$,  behavior discussed in the Appendix~\ref{app:alpha_Scaling_Analysis}. For a fixed $\Delta t$, decreasing $\alpha$ drives the system into the regime $\Delta t \gtrsim \alpha$ (up to a prefactor set by other features of the system; see Eq.~\eqref{eq:crossover_time_L} in Appendix~\ref{app:alpha_Scaling_Analysis})
, where the Strang residual from Step~1, scaling as $|\alpha|\,\Delta t^3$, dominates over the splitting error from Step~2, scaling as $\alpha^2\,\Delta t^2$ (which coincides with the leading-order error of $U_{\text{THRIFT}}^{[1]}$). Their ratio therefore scales as $\Delta t/|\alpha|$, causing the effective exponent of $\alpha$ to cross over from $2$ to $1$, and the advantage of $U_{\text{PS}}^{[1]}$ over $U_{\text{THRIFT}}^{[1]}$ is consequently lost. 

A different behavior is observed for the second-order methods (gray filled squares). In this case, neither method uniformly outperforms the other over the entire range of $\alpha$ values, and the most accurate approximation depends on the interaction strength. Nevertheless, the error ratio clearly approaches unity as $\alpha \rightarrow 0$, indicating that both second-order formulas become asymptotically equivalent in this limit. This behavior is consistent with the theoretical analysis, since both methods exhibit the same leading-order error scaling, $\mathcal{O}(\alpha^2 t^3)$, differing only in higher-order contributions and prefactors. Nevertheless, the convergence toward the asymptotic limit always occurs from below, with $U_{\text{OG}}^{[2]}$ consistently exhibiting a smaller splitting error than $U_{\tilde{\lambda}}^{[2]}$ for sufficiently small values of $\alpha$. For first-order product formulas, the difference between $U_{\text{PS}}^{[1]}$ and $U_{\text{THRIFT}}^{[1]}$ is negligible when $\alpha \to 1$, in correspondence with the scaling with $\alpha$ discussed above in the last section. There is no suppression of the leading error term in this regime, and the two methods become indistinguishable. In contrast, for second-order product formulas, $U_{\text{OG}}^{[2]}$ and $U_{\tilde{\lambda}}^{[2]}$, the behavior outside the regime of small $\alpha$ is less predictable.

\begin{figure}[t]
    \centering
    \includegraphics[width=1\columnwidth]{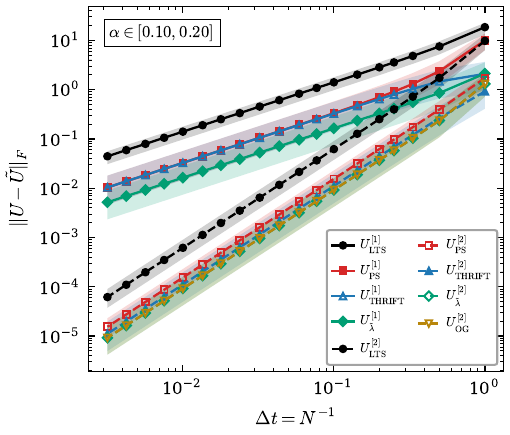}
    \caption{
Propagator error $\|U - \tilde{U}\|_F$ as a function of the time step $\Delta t = N^{-1}$ on a log--log scale, for the $3\times3$ open-boundary two-dimensional spin-$\frac{1}{2}$ XY model [Eq.~\eqref{eq:XY model}] with $\alpha_x = -\alpha_y = \alpha \in [0.10,0.20]$, evaluated at the final time $t = 1$~a.u.
}
    \label{fig:dtsweep}
\end{figure}

Figure~\ref{fig:dtsweep} shows the error as a function of the simulation time step $\Delta t$ (equivalently, the number of Trotter steps) on a log--log scale for the ($3\times3$) open-boundary two-dimensional spin-$\frac{1}{2}$ XY model [Eq.~\eqref{eq:XY model}] with $\alpha_x=-\alpha_y=\alpha = 0.15$, evaluated at the final evolution time $t=1$. As expected, all methods recover their predicted asymptotic scaling in the limit $\Delta t\rightarrow0$: first-order formulas (solid lines) exhibit $\mathcal{O}(\Delta t)$ scaling, whereas second-order formulas (dashed lines) exhibit $\mathcal{O}(\Delta t^2)$ scaling. The shaded region indicates the range of errors obtained over the interval $\alpha\in[0.10,0.20]$, in all cases following the trend of the exemplary $\alpha = 0.15$.

Consistent with the analysis of the previous section, the difference between the first-order Partial--Strang formula, $U_{\text{PS}}^{[1]}$, and the first-order THRIFT formula, $U_{\text{THRIFT}}^{[1]}$, becomes negligible as $\Delta t\rightarrow0$ throughout the parameter range considered. This behavior agrees with the asymptotic convergence derived in Appendix~\ref{app:alpha_Scaling_Analysis}; see Fig.~\ref{fig:alpha_convergence_vs_dt} for a quantitative comparison. Conversely, for the second-order methods (except for $U_{\text{PS}}^{[2]}$), the performance is negligible for all values of $\Delta t$. In these cases, no additional properties beyond those discussed above are required to achieve the $\alpha^2$ scaling of the error. 

A more pronounced distinction emerges outside the asymptotic regime. For larger time steps ($\Delta t\sim1$, or equivalently a small number of steps $N$), $U_{\text{THRIFT}}^{[1]}$ consistently yields smaller absolute errors than both $U_{\text{PS}}^{[1]}$ and $U_{\text{LTS}}^{[1]}$. The same qualitative behavior is observed for the corresponding second-order formulas. 

The results presented in this section indicate that THRIFT provides the highest accuracy among the first-order product formulas considered here. However, when THRIFT cannot be efficiently implemented, $U_{\text{PS}}^{[1]}$ is preferable for first-order schemes. For second-order product formulas, the differences between the methods analyzed in this work are negligible for small $\alpha$. However, as $\alpha \to 1$, both $U_{\text{OG}}^{[2]}$ and $U_{\tilde{\lambda}}^{[2]}$, with the optimized parameter $\tilde{\lambda}$, outperform $U_{\text{THRIFT}}^{[2]}$. The following section presents specific cases in which these product formulas are particularly well suited.

\subsection{Molecular Hamiltonians}
\label{sec:molecular}

The standard workflow for simulating electronic Hamiltonians in second quantization typically proceeds by mapping the fermionic operators to qubit operators and subsequently decomposing the resulting Pauli strings into independently exponentiable terms ~\cite{Langkabel_Bande_QuantumCompute}. The  electronic Hamiltonian in second quantization reads~\cite{Szabo1996}:
\begin{align} \label{eq:molecularHami}
\hat{H} = E_\mathrm{nuc} 
&+ \sum_{pq}\sum_{\sigma} 
h_{pq}\, 
\hat{a}_{p\sigma}^\dagger 
\hat{a}_{q\sigma} \nonumber \\ 
&+ \frac{1}{2}
\sum_{pqrs}
\sum_{\sigma\sigma'}
V_{pqrs}\,
\hat{a}_{p\sigma}^\dagger 
\hat{a}_{q\sigma'}^\dagger 
\hat{a}_{r\sigma'} 
\hat{a}_{s\sigma}.
\end{align}
where $E_\mathrm{nuc}$ is the constant nuclear repulsion energy for a fixed molecular geometry, $\hat{a}_{p\sigma}^\dagger$ and $\hat{a}_{p\sigma}$ are fermionic creation and annihilation operators acting on spatial orbital $p$ with spin $\sigma \in \{\alpha,\beta\}$, $h_{pq}$ are the one-electron integrals encoding kinetic energy and nuclear attraction between spatial orbitals, and $V_{pqrs}$ are the two-electron repulsion integrals (two-body tensor $\equiv V$) defined over spatial orbitals~\cite{Szabo1996}.

The double factorization (DF) framework~\cite{Cohn2021QuantumFilterDF, Sung2026ffsim} allows to implement this procedure at a more abstract, however, structured level for quantum simulations in molecular systems. The electronic Hamiltonian in DF form can be rewritten as:
\begin{align}\label{eq:Molecular_Hamiltonian_Double_Factorization_Form}
    \hat{H} = \text{const.}
    &+ \hat{U}_0^\dagger
    \!\left(
    \sum_{p\sigma} 
    Z^{(0)}_{p\sigma}\, 
    \hat{n}_{p\sigma} 
    \right)\!
    \hat{U}_0
     \nonumber \\ 
     &+ \sum_{k=1}^L 
     \hat{U}_k^{\dagger}
      \!\left(
      \sum_{p\sigma,q\sigma'}
      Z^{(k)}_{p\sigma,q\sigma'} 
      \hat{n}_{p\sigma}
      \hat{n}_{q\sigma'}
      \right)\!
      \hat{U}_k,
\end{align}
Starting from the left, the $\mathrm{const.}$ term collects the nuclear repulsion energy together with the scalar shifts arising from the one- and two-body operators. In the single-particle Hamiltonian, $\hat{n}_{p\sigma}=\hat{a}_{p\sigma}^{\dagger}\hat{a}_{p\sigma}$ denotes the fermionic number operator associated with spatial orbital $p$ and spin $\sigma$. The coefficients $Z_{p\sigma}^{(0)}$ are the eigenvalues of the one-body matrix defined over spin orbitals, obtained by diagonalization with the unitary transformation $\hat{U}_0$. The two-body coefficients $Z_{p\sigma,q\sigma'}^{(k)}$ arise from a low-rank factorization of the two-electron tensor $V_{pqrs}$ defined over spatial orbitals. Since $V_{pqrs}$ is positive semi-definite, it admits a Cholesky or density-fitting decomposition,

\begin{equation*}
V_{pqrs}\approx \tilde{V}_{pqrs}=\sum_kL_{pq}^{(k)}L_{rs}^{(k)}\quad,
\end{equation*}

where $\tilde{V}$ is the approximation of the two-body tensor. Each factor $L^{(k)}$ is diagonalized by a unitary $\hat{U}_k$ to obtain $Z_{p\sigma,q\sigma'}^{(k)}$. Consequently, each contribution $L^{(k)} = \hat{U}_k^{\dagger}\hat{Z}_k\hat{U}_k$, with

\begin{equation}
\hat{Z}_k=\sum_{p\sigma,q\sigma'}Z_{p\sigma,q\sigma'}^{(k)}\,\hat{n}_{p\sigma}\hat{n}_{q\sigma'} \quad \forall k \geq1,
\end{equation}
corresponds to a unitary orbital rotation followed by a diagonal quadratic operator. This structure is particularly well suited for gate-based quantum-circuit implementations, especially for Hamiltonian time evolution~\cite{Cohn2021QuantumFilterDF,Rocca_Cortes_Gonthier_2024}, and is exploited by several fermionic quantum simulation packages such as~\cite{Sung2026ffsim}. Analogously:

\begin{equation}
    \hat{Z}_0 =
    \sum_{p\sigma} 
    Z^{(0)}_{p\sigma}\, 
    \hat{n}_{p\sigma}\quad.  
\end{equation}

If the Hamiltonian additionally satisfies the condition in Eq.~\eqref{eq:thrift_general_hamiltonian}, namely that one leading Hamiltonian $H_0$ (it may be composed of several terms) dominates all other $\alpha_k H_k^{(1)}$ in norm, then the product formulas developed in this work are applicable. 
Eq.~\eqref{eq:Molecular_Hamiltonian_Double_Factorization_Form} exhibits a natural partitioning into a dominant one-body term--used as $H_0$--and a set of two-body fragments--used as perturbative $H_k^{(1)}$. 

The effects of DF and application of the different factorization methodologies is explored by studying the real-time evolution of an initial quantum state under the molecular Hamiltonian for five molecules, H2 and some second-row binary hydrides. They are at their experimental equilibrium geometries, namely H$_2$ (H--H bond length $0.741$~\AA)~\cite{HuberHerzberg1979}, LiH (Li--H bond length $1.595$~\AA)~\cite{HuberHerzberg1979}, BeH$_2$ (linear, Be--H bond length $1.334$~\AA)~\cite{NIST_CCCBDB}, H$_2$O (O--H bond length $0.9572$~\AA, H--O--H angle $104.52^\circ$)~\cite{NIST_CCCBDB}, and NH$_3$ (N--H bond length $1.0124$~\AA, H--N--H angle $106.67^\circ$, $C_{3v}$ symmetry)~\cite{NIST_CCCBDB}. Initially the restricted Hartree--Fock orbitals were obtained and the one- and two-electron integrals generated with \texttt{PySCF}~\cite{PySCF2018}, using a small STO-3G basis set, such that reference calculations are feasible. 

To perform these simulations using the product formulas proposed in Sec.~\ref{sec:methods}, the open-source package/simulation library \texttt{ffsim} was employed \cite{Sung2026ffsim}. Herein, the electronic Hamiltonian was first brought into its DF form and then time-evolved using the built-in implementations of Lie--Trotter--Suzuki product formulas. The methodologies unavailable in \texttt{ffsim} were implemented in this work. \texttt{ffsim} represents the wavefunction directly in the fixed particle-number (full configuration-interaction) sector and applies the product-formula steps at the state vector level, avoiding the explicit construction of the full propagator. This is unlike the simulations presented in Sec.~\ref{sec:XY model}, where the exact and approximate propagators are explicitly available. Consequently, the natural quantity to compare is the state-propagation error, i.e. the $\ell_{2}$ state distance \cite{golub2013}
\begin{equation}
\|(U - \tilde{U})\,|\psi\rangle\|.
\end{equation}

Before turning to the results of the propagation error, the applicability of the product formulas is benchmarked. To this end the DF factorization~\eqref{eq:Molecular_Hamiltonian_Double_Factorization_Form} is exploited regarding the desired smallness of the interaction terms. The target is to compute, whether the one-body term systematically dominates the two-body terms, i.e. 
\begin{equation}
    \|\hat{Z}_0\| > \|\hat{Z}_k\|, \qquad \forall\, k \ge 1.
\end{equation}
This is quantified in Table~\ref{tab:df_benchmark} via the largest ratio $\alpha_k \equiv \|\hat{Z}_k\|/\|\hat{Z}_0\|$ over all $k$. And indeed, empirically it is found that for the molecules studied here, it is true that $\alpha_k << 1$. This difference between the operator norms has direct consequences for the error of the product formulas that are proposed here: Methods that explicitly exploit the smallness of $\alpha_k$ are expected to incur smaller errors than those that do not. Nevertheless, the error bounds of Sec.~\ref{sec:methods} (and Appendix \ref{app:multiple_small_parameters}) are scaling statements at a fixed set of terms: for a given $L$, the error of $U_{\text{PS}}^{[1]}$ scales as $\bar\alpha^2$ compared to $\bar\alpha$ for $U_{\text{LTS}}^{[1]}$, being $\bar{\alpha}$ the average value of $\alpha_k$. For a molecular Hamiltonian this limit cannot be taken: each molecule fixes its own number of two body terms in the double factorization framework (at least the exact number of them, since is related with the number of orbitals), their norm ratios and their commutators, so that what can be compared is the gain at a given Hamiltonian, which is set by these prefactors rather than by $\bar\alpha$ alone. This makes its analysis less direct than for the model systems of Sec.~\ref{sec:XY model}, where $\alpha$ is a free parameter.

\begin{table}[t]
\caption{%
  Double-factorization benchmark of molecules in STO-3G basis. $N_\mathrm{orb}$ is the number of spatial orbitals, $L = N_\mathrm{orb}(N_\mathrm{orb}+1)/2$ that of DF terms (full rank), at which the two-body tensor is reproduced exactly up to the floating-point machine precision ($\|V - \tilde{V}\| \lesssim 10^{-14}$). $\max_k \alpha_k$ is the largest ratio $\alpha_k$ over the two-body terms $k \ge 1$, with $\|\cdot\|$ the Frobenius norm of the operators within the $N$-electron sector of the molecule. $G$ is the estimate of the reduction of the error of $U_{\text{PS}}^{[1]}$ with respect to $U_{\text{LTS}}^{[1]}$ given by Eq.~\eqref{eq:gain_estimate}.
}
\label{tab:df_benchmark}
\begin{ruledtabular}
\begin{tabular}{l c c c c}
Molecule & $N_\mathrm{orb}$ & $L$ & $\max_k \alpha_k$ & $G$ \\
\midrule
H$_2$        &   2 &    3 & 0.50 & 11.5 \\
LiH          &   6 &   21 & 0.27 &  7.7 \\
BeH$_2$      &   7 &   28 & 0.26 &  8.0 \\
H$_2$O       &   7 &   28 & 0.22 & 10.3 \\
NH$_3$       &   8 &   36 & 0.22 &  9.0 \\
\end{tabular}
\end{ruledtabular}
\end{table}

For a given molecule, the gain of $U_{\text{PS}}^{[1]}$ over $U_{\text{LTS}}^{[1]}$ can nevertheless be traced to the structure of the leading error. In one step of $U_{\text{LTS}}^{[1]}$ this error consists of two kinds of commutators: those of the one-body term $\hat{Z}_0$ with each two-body term, and those of the two-body terms among themselves. The symmetric splitting of $U_{\text{PS}}^{[1]}$ removes the first kind exactly choosing a small enough $\Delta t$ at each time step, and keeps the second. The gain therefore depends on how large the two-body--two-body commutators are compared with the one-body--two-body ones, and not on $\bar{\alpha}$ alone. An estimate of this conmutation relation is given by the spectral-range bound of Ref.~\cite{MartinezMartinez2023assessment}:

\begin{align}
  \Big\|\sum_{k\ge1} [\hat{U}_0^{\dagger}\hat{Z}_0 \hat{U}_0,\hat{U}_k^{\dagger}\hat{Z}_k \hat{U}_k]\Big\| &\leq 2\,\|\hat{Z}_0\|\sum_{k\ge1} \|\hat{Z}_k\| , \label{eq:bound_1b2b}\\
  \Big\|\sum_{1\le j<k} [\hat{U}_j^{\dagger}\hat{Z}_j \hat{U}_j,\hat{U}_k^{\dagger}\hat{Z}_k \hat{U}_k]\Big\| &\leq 2\sum_{1\le j<k} \|\hat{Z}_j\|\,\|\hat{Z}_k\| . \label{eq:bound_2b2b}
\end{align}

\begin{figure}[t]
    \includegraphics[width=1\columnwidth]{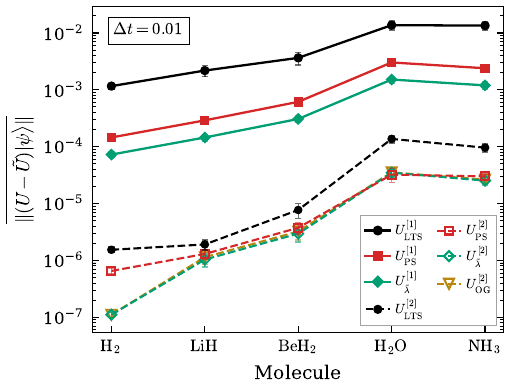}
    \caption{Mean state error $\overline{\|(U - \tilde{U})\,|\psi\rangle\|}$ displaced for the molecules ordered by increasing $N_{orb}$. Solid and dashed lines denote first- and second-order product formulas, respectively. Error bars indicate one standard deviation over ten random initial states.}
    \label{fig:three_panels}
\end{figure}

Eq.~\eqref{eq:bound_1b2b} bounds the part of the error that $U_{\text{PS}}^{[1]}$ removes and Eq.~\eqref{eq:bound_2b2b} the part it keeps. Dividing by the bound for $U_{\text{LTS}}^{[1]}$, which contains both error bounds, by that for $U_{\text{PS}}^{[1]}$ gives an estimate of the reduction, i.e.
\begin{equation}\label{eq:gain_estimate}
  G = \frac{\|\hat{Z}_0\|\sum_{k}\|\hat{Z}_k\| + \sum_{j<k}\|\hat{Z}_j\|\,\|\hat{Z}_k\|}{\sum_{j<k}\|\hat{Z}_j\|\,\|\hat{Z}_k\|} .
\end{equation}
For the molecules of Table~\ref{tab:df_benchmark} $G\approx 8$--$11$. As shown below, this is an upper estimate that reproduces the order of magnitude of the measured gains ($4.5$--$8$, Fig.~\ref{fig:three_panels}), although not their individual values. The same bound explains why the one-body term is the natural choice for the symmetric splitting: having the largest norm and, in general, commuting with none of the two-body terms, $\hat{Z}_0$ carries the dominant share $\|\hat{Z}_0\|\sum_k\|\hat{Z}_k\|$ of the bound. Further research on the structure of the two-body--two-body commutators may allow to improve the estimate of $G$ and therefore the prediction of the gain of $U_{\text{PS}}^{[1]}$ over $U_{\text{LTS}}^{[1]}$ for a given molecule.

Next, the performance of all the approximations is compared in the time-evolutions. To perform these simulations using the product formulas proposed in Sec.~\ref{sec:methods}, the open-source package/simulation library \texttt{ffsim} was employed \cite{Sung2026ffsim}. Herein, the electronic Hamiltonian was first brought into its DF form and then time-evolved using the built-in implementations of Lie--Trotter--Suzuki product formulas. The methodologies unavailable in \texttt{ffsim} were implemented in this work. \texttt{ffsim} represents the wavefunction directly in the fixed particle-number (full configuration-interaction) sector and applies the product-formula steps at the state vector level, avoiding the explicit construction of the full propagator. This is unlike the simulations presented in Sec.~\ref{sec:XY model}, where the exact and approximate propagators are explicitly available. Consequently, the natural quantity to compare is the state-propagation error, i.e. the $\ell_{2}$ state distance \cite{golub2013}
\begin{equation}
\|(U - \tilde{U})\,|\psi\rangle\|.
\end{equation}
The exact reference evolution $U|\psi\rangle$ was obtained with \texttt{scipy.sparse.linalg.expm\_multiply}. To obtain error bars, the equation was evaluated over an ensemble of several propagations starting from randomly generated initial states. Here ten initial states were obtained by applying a Haar-random orbital rotation to the Hartree--Fock determinant~\cite{Sung2026ffsim}; a fixed master seed makes this ensemble reproducible. All time evolution were executed until a final time $t=1$ a.u. using $N=100$ steps ($\Delta t=0.01$ a.u.). The gain expected for the methods proposed in this work is already achieved at this time step as was corroborated in our simulations using smaller time steps, which did not change the results. Figure~\ref{fig:three_panels} reports the mean state distance $\overline{\|(U - \tilde{U})\,|\psi\rangle\|}$ for the molecular systems listed in Table~\ref{tab:df_benchmark}. To say it first, the error bars, computed over the ensemble of $10$ random initial states, are imperceptible on the scale of the figure, confirming that the results are robust with respect to the choice of initial state. The figure shows a clear and consistent reduction in the Lie--Trotter--Suzuki error $\overline{\|(U - U_{\text{LTS}}^{[1]})\,|\psi\rangle\|}$ (black solid line) along with the reduction of the system size across all molecules, which is the largest error found across the methods. The first-order methods $U_{\text{PS}}^{[1]}$ and $U_{\tilde{\lambda}}^{[1]}$ (red and green solid lines) give a smaller error with following the same trend. The second-order methods (dashed lines, same colors) outperform the first-order methods, with the same ordering of methods. Different to $U_{\text{PS}}^{[1]}$, $U_{\text{PS}}^{[2]}$ proves highly competitive with $U_{\tilde{\lambda}}^{[2]}$ and also $U_{\text{OG}}^{[2]}$ (gold dashed line), achieving comparable or lower state-distance errors for the molecules despite its simpler construction.

There are numerous studies in the literature on product formulas for molecular Hamiltonians, including approaches based on the double-factorization framework as well as methods that treat the Hamiltonian directly in its standard representation as a sum of Pauli terms~\cite{Cohn2021QuantumFilterDF, MartinezMartinez2023assessment}. The product formulas proposed in this work have the same implementation requirements as those commonly considered standard in the literature and are therefore equally applicable, provided that the number of individual terms to be evaluated remains computationally feasible and efficient.

\subsection{Circuit Analysis}
\label{app:circuit_analysis}

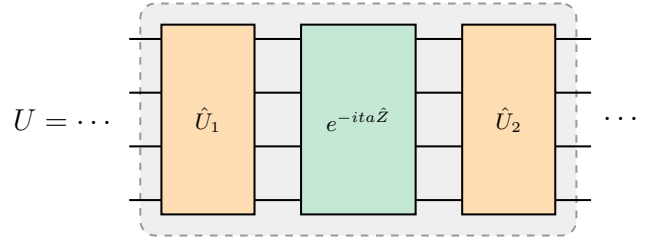
\begin{figure}[t]
\centering
\resizebox{\columnwidth}{!}{%
  \begin{tikzpicture}[
    font=\small,
    tgate/.style  = {draw=black, fill=gateorange!30, thick,
                     minimum width=1.3cm, minimum height=2.65cm,
                     align=center, inner sep=3pt},
    tgateUs/.style= {draw=black, fill=gategreen!30, thick,
                     minimum width=1.6cm, minimum height=2.65cm,
                     align=center, inner sep=2pt},
  ]
  \def\Ybot{0}
  \def\WxL{1.2}
  \def\WxR{7.5}
  \pgfmathsetmacro{\YbotMid}{\Ybot - 1.125}
  \foreach \yy in {0, -0.75, -1.5, -2.25}{
    \draw[thick] (\WxL, \Ybot+\yy) -- (\WxR+0.15, \Ybot+\yy);
  }
  \node[font=\large] at (0.3, \YbotMid) {$U=\cdots$};
  \node[tgate]   (s1) at (2.3,  \YbotMid) {$\hat{U}_1$};
  \node[tgateUs] (s2) at (4.4,  \YbotMid) {$e^{-ita\hat{Z}}$};
  \node[tgate]   (s3) at (6.5,  \YbotMid) {$\hat{U}_2$};
  \node[font=\large] at (8.1, \YbotMid) {$\cdots$};
  \begin{scope}[on background layer]
    \node[fill=gray!15, opacity=0.8, rounded corners=6pt,
          fit=(s1)(s2)(s3), inner sep=8pt] (bbox) {};
    \node[draw=black!40, dashed, thick, rounded corners=6pt,
          fit=(s1)(s2)(s3), inner sep=8pt] (dbox) {};
  \end{scope}
  \node[font=\small\bfseries, anchor=south] at (dbox.north) {};
  \end{tikzpicture}
}
\caption{%
(Color online) Building block for a product formula with a Hamiltonian expressed in the DF framework. The gates are orbital rotations $\hat{U}_{1,2}$ (orange) and diagonal (Z-type) exponentials $e^{-ita \hat{Z}}$ (green).
}
\label{fig:strang_zoom}
\end{figure}

In the DF framework, the propagator approximation written as a product formula can generally be formulated by combining Eqs.~\eqref{eq:product_formula} and \eqref{eq:Molecular_Hamiltonian_Double_Factorization_Form} and omitting the $\mathrm{const.}$ into
\begin{equation}\label{eq:product_formula_UZU}
 U=\prod_j\prod_k \hat{U}_{\pi_j(k)}^{\dagger}e^{-ita_{(j,k)}\hat{Z}_{\pi_j(k)}}\hat{U}_{\pi_j(k)}.
\end{equation}
$\hat{U}_{\pi_j(k)}$ is an orbital rotation and $\hat{Z}_{\pi_j(k)}$ is a diagonal operator. $\hat{U}_{\pi_{j_1}(k_1)} \hat{U}_{\pi_{j_2}(k_2)}^{\dagger}$ is another orbital rotation. The circuit has the general form displayed in Fig.~\ref{fig:strang_zoom}.

The implementation of the diagonal $Z$-type exponential in Eq.~\eqref{eq:product_formula_UZU} follows standard constructions. For a single qubit, $e^{-i\theta Z}$ corresponds to a native $Z$-axis rotation $R_Z(2\theta)$. For multi-qubit operators of the form $e^{-i\theta Z_1 Z_2 \cdots Z_m}$, the exponential can be implemented using a sequence of CNOT gates, followed by a single $R_Z$ rotation and the inverse CNOT sequence, requiring $2(m-1)$ CNOT gates~\cite{NielsenChuang2010}.  

The dominant contribution to the circuit complexity arises from the orbital rotations $\hat{U}_{\pi_j(k)}$, which are decomposed into products of two-level unitary transformations acting on two-dimensional subspaces. Following Ref.~\cite{Oumarou2024}, each orbital rotation is factorized into a sequence of Givens rotations, where each Givens rotation acts on a pair 
of qubits~\cite{GivensRotations2022}. 

\section{Conclusions and Outlook}
\label{sec:conclusions}

First- and second-order product formulas that exploit a hierarchy in the operator norms have been considered, leading to a corresponding error reduction. Unlike the THRIFT method introduced in Ref.~\cite{THRIFT2025} and discussed throughout this paper, the product formulas presented here follow the general structure given in Eq.~\eqref{eq:product_formula}. The only requirement is that they can be implemented efficiently, which is a less restrictive condition.

These benefits were illustrated using a two-dimensional spin-$\frac{1}{2}$ XY model and a set of molecular systems. For the XY model, the errors of the product formulas proposed in this work were shown to scale as $\alpha^2$ and therefore outperform the standard Lie--Trotter--Suzuki product formula, which scales as $\alpha$, and to be competitive with the THRIFT method. Although the molecular Hamiltonians are not parameterized by $\alpha$ and therefore scaling of the error with respect to $\alpha$ is not directly applicable, the product formulas were shown to reduce the state-propagation error by a factor of $4.5$--$8$ compared with the standard Lie--Trotter--Suzuki product formula, depending on the molecule considered. The results of this work indicate that the product formulas proposed here are promising alternatives to the LTS product formulas and widely applicable to a broad range of Hamiltonians, including those that do not satisfy the requirements for THRIFT product formulas. In all these cases, improving the fidelity of the underlying product formula translates into enhanced algorithmic accuracy and more efficient resource usage. 

Many quantum algorithms for example, rely on the efficient implementation of the unitary propagator $U=e^{-iHt}$ associated with a Hamiltonian $H$. A prominent example is Quantum Phase Estimation (QPE). In its standard formulation, QPE requires the implementation of $e^{-iH}$, which corresponds to a time evolution of one atomic unit of time. Reducing the approximation error of the propagator directly will improve the phase estimation accuracy and can potentially lower the overall circuit depth~\cite{Hejazi2024}. More generally, efficient product-formula decompositions are a key ingredient in many quantum algorithms, including variational algorithms based on Hamiltonian-inspired ansätze~\cite{Tilly2022}, digitally simulated adiabatic state preparation~\cite{Hegade2021}, and the evaluation of time-dependent correlation functions~\cite{Cirstoiu2020}. 

Although the Hamiltonians considered in this work are time independent, the methodology developed can be naturally extended to time-dependent systems. In particular, the norm-hierarchy product formulas could be combined with techniques such as the Magnus expansion and time-dependent product formulas~\cite{Ikeda2023minimum} to efficiently simulate time-dependent Hamiltonians, potentially improving the accuracy of dynamical observables while reducing circuit depth. Furthermore, it would be interesting as well to investigate the performance of these product formulas in conjunction with different qubit partitioning strategies~\cite{verteletskyi2020minclique,Yen2023deterministic,MartinezMartinez2023assessment}.

\section{Code Availability}\label{sc:Code_availability}
The code used to produce the results and figures in this work is available at \url{https://github.com/CarlosPineiroMoreno/norm-aware-product-formulas}. The repository is currently private and will be made public upon publication; access can be requested from the corresponding author in the meantime.

\begin{acknowledgments}
The authors are grateful for the funding of this research via the Lower Saxony Profile Professorship jointly issued by the Ministry of Science and Culture of Lower Saxony and the Volkswagen Foundation as well as from the Deutsche Forschungsgemeinschaft (DFG) under Germany’s Excellence Strategy within the Cluster of Excellence PhoenixD (Photonics, Optics, and Engineering -- Innovation Across Disciplines, EXC 2122, projectID 390833453) for funding of the position of C.A.P.M.
\end{acknowledgments}

\appendix

\bibliographystyle{apsrev4-1}
\bibliography{references}

\section{Proof of Lemma~\ref{lem:lambda_decomp}}\label{app:lambda_proof}

Let us define
\begin{align}\label{eq:S_lambda_decomp}
  S(t)=&\,e^{-i \lambda H_0 t}
  \, e^{-i \alpha H^{(1)} t/2}
  \, e^{-i (1-2\lambda) H_0 t}
  \nonumber\\
  &\times e^{-i \alpha H^{(1)} t/2}
  \, e^{-i \lambda H_0 t} \quad,
\end{align}

$S(t)$ trivially satisfies the identity

\begin{equation}
    S(t)S(-t)=\mathbf{1} \quad,
\end{equation}

where $\mathbf{1}$ is the identity operator. Due to the argument of Yoshida in~\cite{Yoshida1990}, $S(t)$
admits an expansion of the form

\begin{equation}\label{eq:yoshida_S}
    S(t) = e^{(x_1t+x_3t^3+x_5t^5+\cdots)}\;.
\end{equation}

Since $S(t)$ is symmetric by construction, only odd powers of $t$ appear in the exponent. To determine the order of $S(t)$, the coefficients $x_1$ and $x_3$ are explicitly derived via repeated application of the Baker--Campbell--Hausdorff formula (BCH)~\cite{Hall2015LieGroups} for the explicit product of exponential terms of Eq. \eqref{eq:S_lambda_decomp}, giving

\begin{align}
&e^{-i \alpha H^{(1)} t/2}
\, e^{-i (1-2\lambda) H_0 t}
\, e^{-i \alpha H^{(1)} t/2}
 =
\exp\!\Biggl(
- i t \bigl[
\alpha H^{(1)}
+\nonumber\\
&
(1-2\lambda) H_0
\bigr]+
t^{3} \mathcal{E}_\lambda
+
\mathcal{O}(t^{5})
\Biggr),
\end{align}

where 

\begin{align}
\mathcal{E}_\lambda
&=
-\frac{i}{24}
\, \alpha^{2} (1-2\lambda)
\,[H^{(1)},[H^{(1)},H_0]]
\nonumber\\
&\quad +
\frac{i}{12}
\, \alpha (1-2\lambda)^{2}
\,[H_0,[H_0,H^{(1)}]].
\end{align}

With that $S(t)$ becomes

\begin{align}
S(t) &= e^{-i \lambda H_0 t}
\exp\Bigl(
- i t \bigl[\alpha H^{(1)}+(1-2\lambda) H_0\bigr]
\nonumber\\
&\quad + t^3 \mathcal{E}_\lambda
+ \mathcal{O}(t^5)
\Bigr)
e^{-i \lambda H_0 t}\nonumber\\
&=
\exp\Bigl(
- i t (H_0 + \alpha H^{(1)})+ \alpha t^3
\bigl(
\gamma_1 [H^{(1)},[H^{(1)},H_0]]
\nonumber\\
&\quad
+ \gamma_2 [H_0,[H_0,H^{(1)}]]
\bigr)
+ \mathcal{O}(t^5)
\Bigr),
\end{align}

where, by comparison with Eq. \eqref{eq:yoshida_S}, the variables are found to be

\begin{align}
    x_1 &=-iH=-i(H_0+\alpha H^{(1)}) \\
    x_3 &= i\alpha
\bigl(
\gamma_1 [H^{(1)},[H^{(1)},H_0]]
+ \gamma_2 [H_0,[H_0,H^{(1)}]]
\bigr).
\end{align}

Since $x_1 = -iH$, the product formula reproduces the exact generator to leading order in $t$. The first correction appears at order $t^3$ through $x_3$, which confirms that $S(t)$ is a second-order product formula for all values of $\lambda$. The explicit forms of the coefficients $\gamma_1$ and $\gamma_2$ are given in Eq.~\eqref{eq:gamma_coeffs}.

\subsection{Optimal \texorpdfstring{$\lambda$}{lambda}}

The optimal value of $\lambda$ is obtained by minimizing the combined error, $\gamma_1^2 + \gamma_2^2$, where both coefficients depend on $\lambda$ as given in Eq.~\eqref{eq:gamma_coeffs}. It is 
\begin{align}\label{eq:error_polynomial}
  \gamma_1^2 + \gamma_2^2
  &= \alpha^2\!\left(\frac{1-6\lambda}{24}\right)^{\!2}
     +\left(-\frac{1}{12}+\frac{\lambda}{2}-\frac{\lambda^2}{2}\right)^{\!2}.
\end{align}
The minimizing condition $\partial(\gamma_1^2 + \gamma_2^2)/\partial\lambda = 0$ leads to the following cubic equation
\begin{align}\label{eq:cubic_condition}
  f(\lambda,\alpha)
  \equiv
  \left(\lambda - \tfrac{1}{2}\right)
  \left(\lambda^2 - \lambda + \tfrac{1}{6}\right)
+ \frac{\alpha^2}{48}(6\lambda - 1)
  = 0.
\end{align}

In principle, this cubic equation can be solved analytically using standard methods to obtain its three solutions explicitly as functions of $\alpha$. However, rather than analyzing the full analytic structure of these solutions and the nature of the corresponding extrema, the focus will turn into the regime of small $\alpha$, which is the relevant limit for the present analysis. 

Accordingly, the roots are determined perturbatively. As a first step, the unperturbed case $\alpha = 0$ is considered, for which Eq.~\eqref{eq:cubic_condition} reduces to a factorizable polynomial whose solutions are given by

\begin{align}\label{eq:roots_alpha0}
  \lambda_1(0) &= \tfrac{1}{2} - \tfrac{\sqrt{3}}{6}, \quad\quad
  \lambda_2(0) = \tfrac{1}{2} + \tfrac{\sqrt{3}}{6}, \nonumber\\
  \lambda_3(0) &= \tfrac{1}{2}.
\end{align}

By the implicit function theorem~\cite{Rudin1976}, one finds that the leading correction to the above extrema is of order $\alpha^2$.
\begin{align}
  \lambda_1(\alpha) &\approx \!\left(\tfrac{1}{2}-\tfrac{\sqrt{3}}{6}\right)
                              - \tfrac{1}{8}(2-\sqrt{3})\,\alpha^2,
                              \label{eq:root1_expansion}\\
  \lambda_2(\alpha) &\approx \!\left(\tfrac{1}{2}+\tfrac{\sqrt{3}}{6}\right)
                              - \tfrac{1}{8}(2+\sqrt{3})\,\alpha^2,
                              \label{eq:root2_expansion}\\
  \lambda_3(\alpha) &\approx \tfrac{1}{2} + \tfrac{1}{2}\,\alpha^2.
                              \label{eq:root3_expansion}
\end{align}

The next step consists of analyzing the nature of these critical points. Evaluating $\partial f/\partial\lambda$ at each root reveals
\begin{enumerate}[label=(\roman*)]
  \item $\lambda_1(\alpha)$ and $\lambda_2(\alpha)$ are \textit{minima} for small $\alpha$.
  \item $\lambda_3(\alpha) = 1/2$ is a \textit{local maximum} for small $\alpha$.
\end{enumerate}

This could have been anticipated by inspection of Eq.\eqref{eq:error_polynomial}, since for $\alpha=0$, the minima of $\gamma_1^2+\gamma_2^2$ are just the zeros of the polynomial $\left(-\frac{1}{12}+\frac{\lambda}{2}-\frac{\lambda^2}{2}\right)$ which are $\lambda_1(0)$ and $\lambda_2(0)$ introduced above.

Comparing the coefficients of $\alpha^2$ in
\eqref{eq:root1_expansion}--\eqref{eq:root2_expansion}:
\[
  \tfrac{1}{8}(2-\sqrt{3}) \approx 0.0335
  \quad\text{vs}\quad
  \tfrac{1}{8}(2+\sqrt{3}) \approx 0.4665,
\]
$\lambda_1(\alpha)$ is considerably more stable than $\lambda_2(\alpha)$ with respect to changes in $\alpha$ due to its smaller coefficient. In particular, its zeroth-order value, $\lambda_1(0)=\tilde{\lambda} = \tfrac{1}{2} - \tfrac{\sqrt{3}}{6} \approx
0.211325$, was selected as a fixed parameter in the simulations carried out throughout this work for all product formulas depending on $\lambda$. This choice is theoretically motivated by the perturbative stability of this value and was further validated numerically, as it consistently yielded results closest to the expected scaling in $\alpha$ across the range of operating conditions considered. 

To assess the robustness of this model-independent behavior beyond a single class of Hamiltonians, we numerically evaluated the optimal parameter $\arg\min_\lambda \|U-U_{\text{OG}}^{[2]}(\lambda)\|_F$ as a function of the tunable coupling strength $\alpha$ across a broad and diverse set of benchmark systems which encompass a wide range of interaction structures, dimensionalities, and connectivity patterns, including paradigmatic systems extensively studied in quantum many-body physics. For all the systems considered, the optimal $\arg\min_\lambda \|U-U_{\text{OG}}^{[2]}(\lambda)\|_F$
collapses onto the common value $\tilde{\lambda}=\tfrac{1}{2}-\tfrac{\sqrt{3}}{6}$ as $\alpha\to0$. The universality study of $\tilde{\lambda}$ is addressed in the Supplementary Materials.
\section{Supplementary Analysis of the \texorpdfstring{$\alpha$}{alpha}-Scaling}
\label{app:alpha_Scaling_Analysis}
\begin{figure*}[t]
    \centering
    \includegraphics{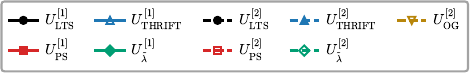}\\[2pt]
    \begin{subfigure}{0.49\textwidth}
        \includegraphics[width=\textwidth]{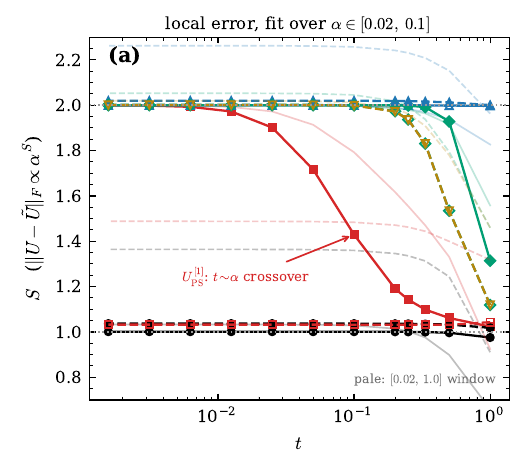}
    \end{subfigure}
    \hfill
    \begin{subfigure}{0.49\textwidth}
        \includegraphics[width=\textwidth]{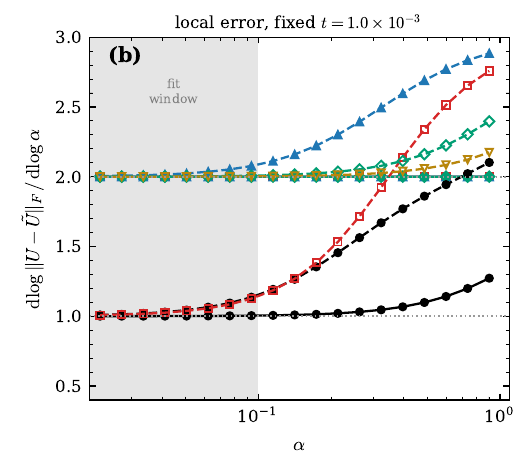}
    \end{subfigure}
    \caption{(a) Power-law exponent $S$ ($\alpha^S$) as a function of the one single time step $t$ (local error) for the different integration methods. Solid lines correspond to first-order methods and dashed lines to second-order methods. The lighter curves in the background indicate a wider range for the fitting. (b) Local exponent $\mathrm{d}\log\|U-\tilde U\|_F/\mathrm{d}\log\alpha$ as a function of $\alpha$ for a single time step $t=10^{-3}$; the shaded region marks the fit window $\alpha\in[0.02,0.1]$ used in (a).}
    \label{fig:alpha_convergence_vs_dt}
\end{figure*}

To further investigate the $\alpha$-scaling, we analyzed the local error of the product formulas as a function of $\alpha$, using the same system as considered in Section~\ref{sec:XY model}. Recalling all the discussion made in the main text, the local error is expected to scale as $\|U-\tilde U(t)\|_F=\mathcal{O}(\alpha^S)$, where the effective exponent $S$ is predicted to transition from $2$ to $1$ for some product formulas as the evolution time $t$ increases. An estimate of the crossover time $t_c$ is derived, at which the effective exponent $S$ transitions from $2$ to $1$. The local error of $U_{\text{PS}}^{[1]}$ is bounded, as a function of $\alpha$ and $t$, as:

\begin{equation}
    \bigl\|U(t)-U_{\text{PS}}^{[1]}(t)\bigr\| \le \alpha^2 t^2 E_1 + |\alpha|\, t^3 E_2 + \mathcal{O}(\alpha^2 t^3) \quad,
\end{equation}
where $E_1$ and $E_2$ are given in Eqs.~\eqref{eq:trotter_alpha_error} and \eqref{eq:strang_local_error}, respectively. It is desired that the second term, concerning the scaling linear in $\alpha$, is much smaller than the first term, which scales quadratically in $\alpha$. When the ratio of the two terms is of the same order, the crossover from the $\alpha^2$ scaling to the linear $\alpha$ scaling is expected to occur. This condition can be expressed as 

\begin{equation}\label{eq:crossover_time}
   \dfrac{|\alpha|\, t_c^3 E_2}{\alpha^2 t_c^2 E_1} \approx 1
   \quad, \qquad
   t_c \approx |\alpha| \dfrac{E_1}{E_2} \quad,
\end{equation}
so that the $\alpha^2$ scaling ($S\simeq2$) holds for $t  \lesssim t_c$, the linear scaling ($S\simeq1$) for $t \gtrsim t_c$, and the crossover takes place at $t \sim t_c$. Since $E_1$ and $E_2$ are upper bounds, $t_c$ is an estimate of the crossover time rather than its exact value. The magnitudes $E_1$ and $E_2$ depend on the system features such as the number of individual terms in the Hamiltonian, their norms and the commutators between them. Considering commutator relations and the norm bound proposed in Ref.~\cite{MartinezMartinez2023assessment} one finds that:

\begin{equation}\label{eq:crossover_time_L}
   t_c \sim |\alpha| L \quad,
\end{equation}
considering the norm of the operators $\|H_k \|=1$ for $k \geq 0$ (the crossover time is rescaled by $\|H_0 \|$ when this operator has a different norm). As shown in Fig.~\ref{fig:alpha_convergence_vs_dt}(a), the behavior predicted is indeed observed. Specifically, for each product formula and for every value of $t$ within the selected time interval, the local error between the product-formula approximation and the exact evolution was computed over two ranges of $\alpha$ indicated in the panel (fit window). The first range, indicated by the solid curves in the figure, corresponds to $\alpha\in[0.02,0.1]$. This range is chosen from the Figure \ref{fig:alpha_convergence_vs_dt}(b) as the region where the local error exhibits a clear power-law behavior in $\alpha$ for a small time step $t$. Figure~\ref{fig:alpha_convergence_vs_dt}(b) therefore illustrates how the local scaling exponent in $\|U-\tilde U(t)\|_F=\mathcal{O}(\alpha^S)$ evolves with $\alpha$ for a fixed time $t$. The fact that the extracted exponent is not constant over the entire fitting window but actually increases for some of the curves when $\alpha \to1$, indicates that the local error is not described by a single power law, but rather by a superposition of different powers of $\alpha$, greater than the predicted asymptotic exponent. Nevertheless, for small values of $\alpha$ the power-law behavior is recovered, and the effective exponent $S$ can be extracted from a log--log fit of the local error as a function of $\alpha$. The second range, indicated by the lighter curves in the background, corresponds to $\alpha\in[0.02,0.2]$. This range is chosen to illustrate how the effective exponent $S$ changes when the fitting window is extended to larger values of $\alpha$, and therefore, other power laws in $\alpha$ contribute to the fitting. In this specific example, for $\alpha \lesssim 5\times10^{-2}$ all curves converge to their predicted asymptotic exponent, within $0.05$. This agrees with the crossover time predicted in \eqref{eq:crossover_time_L} and with the fit window employed.

\section{Multiple small parameters.}\label{app:multiple_small_parameters}
For a Hamiltonian 
\begin{equation}\label{eq:thrift_general_hamiltonian}
    H = H_0 + \sum_{k=1}^{L} \alpha_k H_k^{(1)} \quad \text{with} \quad |\alpha_k| < 1 \quad \forall k ,
\end{equation}
with multiple small parameters $\alpha_k$ the error proportional $t^2$ scales, in analogy to Eq.~\eqref{eq:trotter_alpha_error}, as:

\begin{align}\label{eq:trotter_alphas_error}
  &\Bigl\|
    e^{-i \sum_{k=1}^{L} \alpha_k H^{(1)}_k t}
    - \prod_{k=1}^{L} e^{-i \alpha_k H_k^{(1)} t}
  \Bigr\|
  \nonumber\\
  &\quad \le
  \frac{t^2}{2}
  \sum_{1 \le j < k \le L} |\alpha_j \alpha_k|
  \bigl\| [H_j^{(1)}, H_k^{(1)}] \bigr\|.
\end{align}

To compare with the first-order Lie--Trotter--Suzuki scaling in this situation of having individual coefficients $\alpha_k$, the boundary of the PS error is expressed in terms of the mean absolute value of those coefficients $\overline{\alpha} = \frac{1}{L}\sum_{k=1}^L |\alpha_k|$. Using 

\begin{equation*}
    \sum_{1 \le j < k \le L} |\alpha_j \alpha_k| \le 
\frac{1}{2}\left(\sum_{k=1}^L |\alpha_i|\right)^2 \quad,
\end{equation*}

the worst-case bound of \eqref{eq:trotter_alphas_error} scales as $O\!\left(\overline{\alpha}^2 t^2\right)$, which corresponds to a suppression of the leading error term by a 
factor of $\overline{\alpha}<1$ at a fixed number of terms $L$ with respect to the expected $O(\overline{\alpha}\, t^2)$ 
scaling of the standard first-order Lie--Trotter--Suzuki decomposition; the prefactors contain $L$ and the commutators between the terms. 

To assess the robustness of the $\lambda$ value employed on the main work for the second-order product formulas, we numerically evaluated the optimal parameter $\arg\min_\lambda \|U-U_{OG}^{[2]}(\lambda)\|_F$ as a function of the tunable coupling strength $\alpha$ across a broad and diverse set of benchmark systems which encompass a wide range of interaction structures, dimensionalities, and connectivity patterns, including paradigmatic systems extensively studied in quantum many-body physics. The same search was performed using the $U_{\lambda}^{[2]}$ product formula but in general no different results were obtained in the regime of small $\alpha$. The considered models include one-dimensional spin chains, namely the $XY$ model with open and periodic boundary conditions, the $XY$ model with a transverse field, and the transverse-field Ising and $XXZ$ models with open boundary conditions, as well as higher-connectivity geometries such as a two-dimensional $3\times3$ $XY$ lattice, a $2\times4$ $XY$ ladder, and an all-to-all interacting spin model. A random $k$-local spin model is also included. In addition, we considered a one-dimensional Fermi--Hubbard Hamiltonian, where $\alpha$ is identified with the on-site interaction strength $U$~[46–51] (not to be confused with the exact propagator introduced later). Using the notation of the main text, the Hamiltonians have the general form

\begin{equation}
    H = H_0 + \alpha H^{(1)}  \qquad H^{(1)}=\sum_{k=1}^{L}\,H^{(1)}_k
\end{equation}
where $L$ is the number of terms in $H^{(1)}$. The specific components for each case are detailed in Table~\ref{tab:systems} as well as the interaction graph, that is, how the connections/interactions between different spins in the system were considered.  For the Fermi--Hubbard model, $H_0$ is the hopping term: $H_{0}=-\sum_{\langle i,j\rangle,\sigma}
\left(a_{i\sigma}^{\dagger}a_{j\sigma}
+a_{j\sigma}^{\dagger}a_{i\sigma}\right)$, being $a_{i\sigma}^{\dagger}$ ($a_{i\sigma}$) the creation (anihilation) operator which create (anihilate) a fermion in the site $i$ with spin $\sigma = \{ \uparrow, \downarrow\}$, while  $H^{(1)} = \sum_i n_{i\uparrow}n_{i\downarrow}$, where $n_{i\sigma}=a_{i\sigma}^{\dagger} a_{i\sigma}$ is the fermionic number operator. For the remaining systems we take $H_0=\sum_m Z_m$. The random $k$-local model consists of $L=12$ random Pauli strings with  two or three (randomly selected) on eight qubits non-trivial operations per operator, with standard normal coefficients drawn from a fixed seed and rescaled. In all cases, $\alpha$ represents a genuine physical coupling controlling the relative weight of non-commuting Hamiltonian contributions.

\begin{table*}[t]
    \centering
    \caption{Benchmark systems used in the universality study. All systems share $H = H_0 + \alpha\,H^{(1)}$. For the spin models $H_0=\sum_m Z_m$ and $H^{(1)}$ is a sum over the bonds $\langle ij\rangle$ of the interaction graph. For the Fermi--Hubbard model, 4 electrons (2$\uparrow$+2$\downarrow$) in 5 sites were considered.}
    \label{tab:systems}
    \small
    \begin{tabular}{|l|c|l|l|}
        \hline
        System & System size & Interaction Features & $H^{(1)}$ \\
        \hline
        Chain $XY$ & 8 qubits & open chain (7 bonds) & $\sum_{\langle ij\rangle}(X_iX_j - Y_iY_j)$ \\
        Chain Ising & 8 qubits & open chain & $\sum_{\langle ij\rangle}X_iX_j$ \\
        Chain $XXZ$ & 8 qubits & open chain & $\sum_{\langle ij\rangle}(X_iX_j - Y_iY_j + 0.5\,Z_iZ_j)$ \\
        Ring $XY$ & 8 qubits & periodic chain (8 bonds) & $\sum_{\langle ij\rangle}(X_iX_j - Y_iY_j)$ \\
        Square $3\times3$ $XY$ & 9 qubits & open $3\times3$ lattice (12 bonds) & $\sum_{\langle ij\rangle}(X_iX_j - Y_iY_j)$ \\
        Ladder $2\times4$ $XY$ & 8 qubits & open $2\times4$ lattice (10 bonds) & $\sum_{\langle ij\rangle}(X_iX_j - Y_iY_j)$ \\
        All-to-all & 7 qubits & 21 bonds & $\sum_{i<j}(X_iX_j - Y_iY_j)$ \\
        $XY$ + transverse field & 8 qubits & open chain & $\sum_{\langle ij\rangle}(X_iX_j - Y_iY_j) + 0.7\sum_i X_i$ \\
        Fermi--Hubbard 1D & 10 qubits (5 sites) & open chain, 2$\uparrow$+2$\downarrow$ & $\sum_i n_{i\uparrow}n_{i\downarrow}$ \\
        Random $k$-local & 8 qubits & \begin{tabular}[t]{@{}l@{}}$L=12$ $P_k$ \\ random Gaussian coefficients $c_k$\end{tabular} & $\sum_{k=1}^{L} c_k P_k$ \\
        \hline
    \end{tabular}
\end{table*}

\begin{figure*}[t]
    \centering
    \begin{subfigure}{0.49\textwidth}
        \includegraphics[width=\textwidth]{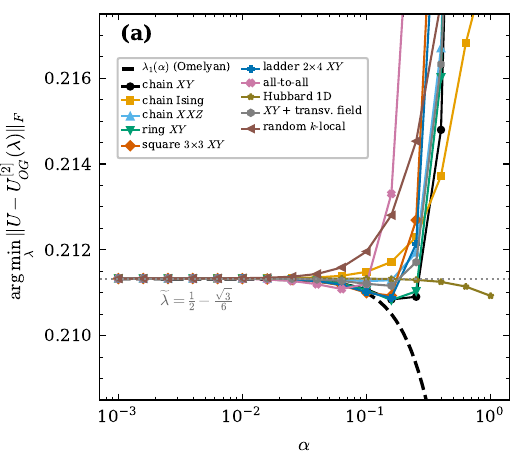}
    \end{subfigure}
    \hfill
    \begin{subfigure}{0.49\textwidth}
        \includegraphics[width=\textwidth]{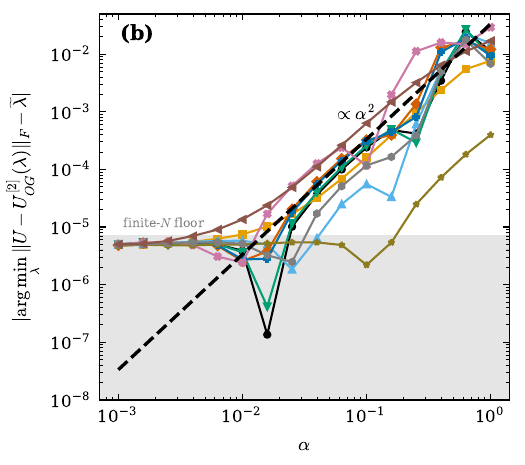}
    \end{subfigure}
    \caption{Universality of the optimal parameter $\tilde{\lambda}$. Data for hand-designed spin systems (see text)
and a $1$D Fermi--Hubbard model.}
    \label{fig:lambda_universality}
\end{figure*}

To evaluate $\arg\min_\lambda \|U-U_{OG}^{[2]}(\lambda)\|_F$ the exact solution $U$ for a time $t=1$ for all the models was found. The product formula  $ U_{OG}^{[2]}(\lambda)$ was computed using $N=100$ time steps. The optimal value $\arg\min_\lambda \|U-U_{OG}^{[2]}(\lambda)\|_F$ was obtained by minimizing the Frobenius norm with the \texttt{scipy.optimize.minimize\_scalar} function of python (see code availability statement). As shown in Fig.~\ref{fig:lambda_universality}(a), the optimal $\lambda$ value $\arg\min_\lambda \|U-U_{OG}^{[2]}(\lambda)\|_F$
collapses onto the common value $\tilde{\lambda}=\tfrac{1}{2}-\tfrac{\sqrt{3}}{6}$ as
$\alpha\to0$ for all systems, irrespective of dimensionality, connectivity, or Hilbert-space
size, but rather diverges when $\alpha$ increases (in a model-dependent way). Panel (b) quantifies the approach: the deviation
$|\arg\min_\lambda \|U-U_{OG}^{[2]}(\lambda)\|_F-\tilde{\lambda}|$ vanishes as $\propto\alpha^{2}$ until it reaches
a floor set by the finite number of time steps $N$ ($\sim\Delta t^{2}$) confirming the asymptotic approach detailed in the Appendix A of the main text. In both graphs were include the minimum of the Eq. (A9) of the Appendix A of the main text as $\lambda_1(\alpha)$ (Omelyan).

\end{document}